\documentclass[12pt]{article}%
\usepackage{cite}
\usepackage{fix-cm}
\usepackage{color}
\usepackage[usenames,dvipsnames]{xcolor}
\usepackage{heck}
\usepackage{amsmath}
\usepackage{amsfonts}
\usepackage{amssymb}
\usepackage{graphicx}
\usepackage{multicol}
\usepackage{hyperref}
\usepackage{setspace}
\usepackage[all]{xy}
\usepackage{verbatim}
\usepackage{color}
\usepackage{epsfig}
\usepackage[margin=1in]{geometry}%
\usepackage{algorithm}
\usepackage{algorithmic}
\providecommand{\U}[1]{\protect\rule{.1in}{.1in}}
\numberwithin{equation}{section}

\def\be{\begin{equation}}
\def\ee{\end{equation}}
\def\bea{\begin{eqnarray}}
\def\eea{\end{eqnarray}}

\definecolor{gray}{rgb}{0.5,0.5,0.5}
\begin{document}

\date{May 2016}

\title{MCMC with Strings and Branes:\\[2mm] The Suburban Algorithm (Extended Version)}

\institution{PENN}{\centerline{${}^{1}$Department of Physics and Astronomy, University of Pennsylvania, Philadelphia, PA 19104, USA}}

\institution{UNC}{\centerline{${}^{2}$Department of Physics, University of North Carolina, Chapel Hill, NC 27599, USA}}

\institution{Analog}{\centerline{${}^{3}$Analog Devices $\vert$ Lyric Labs, One Broadway, Cambridge, MA 02142, USA}}

\institution{Gamalon}{\centerline{${}^{4}$Gamalon Labs, One Broadway, Cambridge, MA 02142, USA}}

\authors{Jonathan J. Heckman\worksat{\PENN, \UNC}\footnote{e-mail: {\tt jheckman@sas.upenn.edu}},
Jeffrey G. Bernstein\worksat{\Analog}\footnote{e-mail: {\tt jeff.bernstein@analog.com}},
Ben Vigoda\worksat{\Gamalon}\footnote{e-mail: {\tt ben.vigoda@gamalon.com}}}

\abstract{Motivated by the physics of strings and branes, we develop a class of Markov chain Monte Carlo (MCMC) algorithms involving extended objects. Starting from a collection of parallel Metropolis-Hastings (MH) samplers, we place them on an auxiliary grid, and couple them together via nearest neighbor interactions. This leads to a class of ``suburban samplers'' (i.e., spread out Metropolis). Coupling the samplers in this way modifies the mixing rate and speed of convergence for the Markov chain, and can in many cases allow a sampler to more easily overcome free energy barriers in a target distribution. We test these general theoretical considerations by performing several numerical experiments. For suburban samplers with a fluctuating grid topology, performance is strongly correlated with the average number of neighbors. Increasing the average number of neighbors above zero initially leads to an increase in performance, though there is a critical connectivity with effective dimension $d_{\mathrm{eff}} \sim 1$,
above which ``groupthink'' takes over, and the performance of the sampler declines.}

\maketitle

\tableofcontents

\enlargethispage{\baselineskip}

\setcounter{tocdepth}{2}

\newpage

\section{Introduction}

Markov chain Monte Carlo (MCMC) methods are a remarkably robust way to sample
from complex probability distributions. In this class of algorithms, the Metropolis-Hastings (MH)
algorithm \cite{Metropolis:1953am, Hastings:1970} stands out as an important
benchmark.

One of the appealing features of the original Metropolis algorithm is the
simple physical picture which underlies the general method. Roughly
speaking, the idea is that the thermal fluctuations of a particle moving in
an energy landscape provides a conceptually elegant way to sample from a target distribution.
Recall that for $X$, a continuous random variable with outcome $x$, we have a
probability density $\pi(x)$, and a proposal kernel $q(x^{\prime} | x)$.
In the MH algorithm, a new value $x^{\text{new}}$ is drawn from
the distribution $q$ and is then accepted with probability:
\begin{equation}
a\left( x^{\text{new}} | x^{\text{old}} \right) = \min\left(
1,\frac{q(x^{\text{old}}|x^{\text{new}})}{q(x^{\text{new}}|x^{\text{old}})}
\frac{\pi(x^{\text{new}})}{\pi(x^{\text{old}})}\right)  .
\end{equation}

On the other hand, there are also well known drawbacks to MCMC methods.
For example, though in many cases there is
an expectation that sampling will converge to the correct posterior
distribution, the actual speed at which this can occur is often unknown. Along
these lines, it is possible for a sampler to remain trapped in a metastable
equilibrium for a long period of time. A related concern is that once a
sampler becomes trapped, a large free energy barrier can obstruct an accurate determination of
the global structure of the distribution. Some of these issues can be overcome by
sufficient tuning of the proposal kernel, or by comparing the performance of
different samplers. It is therefore natural to ask whether further inspiration
from physics can lead to new examples of samplers.

\begin{figure}[ptb]
\centering
\includegraphics[
height=2.4837in,
width=4.4088in
]{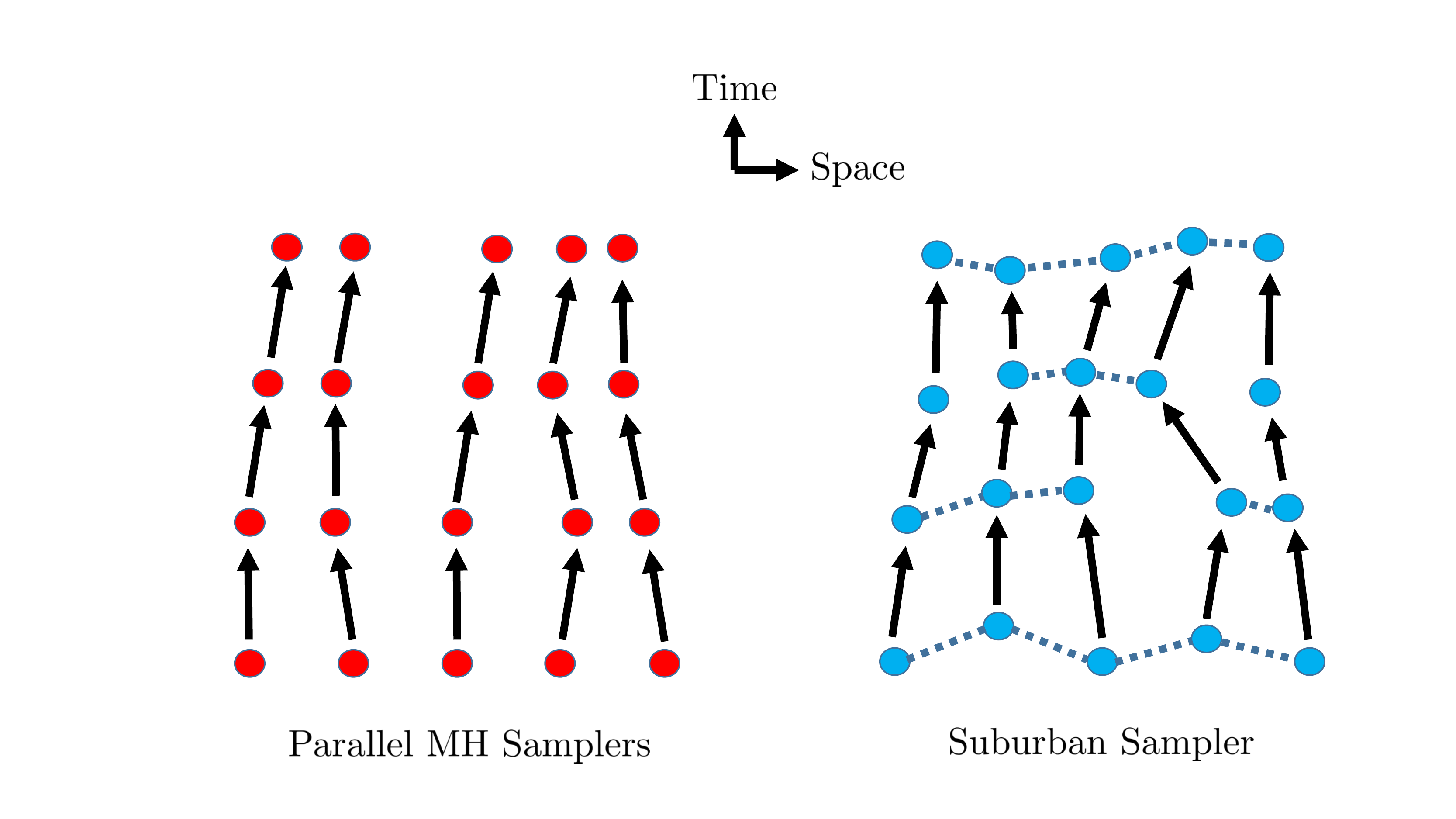}\caption{Depiction of how parallel MH\ samplers (left) and
a suburban sampler (right) evolve as a function of time. In the suburban sampler,
nearest neighbors on a grid can have correlated inferences (depicted by dashed
lines), leading to faster mixing rates. The absence of a dashed line in a given
time step indicates a splitting of the extended object.}%
\label{parstringbrane}%
\end{figure}

Now, although the physics of point particles underlies much of our modern
understanding of natural phenomena, it has proven fruitful, especially in the
context of high energy theoretical physics, to consider objects such as strings and more
generally $p$-branes with finite extent in $p$ spatial dimensions (a string being
a case of a $1$-brane). One of the main features of branes is
that the number of spatial dimensions strongly affects how a localized perturbation
propagates across its worldvolume. Viewing a brane as a collective of point
particles that interact with one another (see figure \ref{parstringbrane}),
this suggests applications to questions in statistical inference \cite{Heckman:2013kza}.

Motivated by these physical considerations, our aim in this work will be to
study generalizations of the MH algorithm for such extended
objects. For an ensemble of $M$ parallel MH samplers of
$\pi(x)$, we can alternatively view this as a single particle sampling from
$M$ variables $x_{1},...,x_{M}$ with density:
\begin{equation}
\pi(x_{1},...,x_{M})=\pi(x_{1})...\pi(x_{M}),
\end{equation}
where the proposal kernel is simply:
\begin{equation}
q_{\text{parallel}}(x_{1}^{\text{new}},...,x_{M}^{\text{new}}|x_{1}%
^{\text{old}},...,x_{M}^{\text{old}})=q(x_{1}^{\text{new}}|x_{1}^{\text{old}%
})...q(x_{M}^{\text{new}}|x_{M}^{\text{old}}).
\end{equation}
To realize an MCMC algorithm for an extended object, we shall keep the same
target $\pi(x_{1},...,x_{M})$, but we will now change the proposal kernel by
interpreting the index $\sigma$ on $x_{\sigma}$ as specifying the location of a
statistical agent in a network. Depending on the connectivity of this network,
an agent may interact with several neighboring agents (if each agent
communicates with no neighbors, this is equivalent to parallel MH samplers).
Schematically then, MCMC with an extended object
involves modifying the proposal kernel to the form:
\begin{equation}
q_{\text{extend}}(x_{1}^{\text{new}},...,x_{M}^{\text{new}}|x_{1}^{\text{old}%
},...,x_{M}^{\text{old}})=\prod_{\sigma=1}^{M}q_{\sigma}(x_{\sigma
}^{\text{new}}|\mathrm{Neighbors\,of\,}x_{\sigma}^{\text{old}}%
).\label{Qextend}%
\end{equation}
In the above, the connectivity of the extended object specifies its overall
topology. For example, in the case of a string, i.e., a one-dimensional
extended object, the neighbors of $x_{i}$ are $x_{i-1}$, $x_{i}$, and
$x_{i+1}$. Figure \ref{parstringbrane} depicts
the time evolution of parallel MH samplers compared with the
suburban sampler.

From this perspective, the suburban algorithm is a particular choice of ensemble MCMC.
Ensemble samplers have been considered previously in the MCMC
literature (see e.g., \cite{SwendsenWang, Geyer, AdaptiveDirectSamp,
ParallelTempering,  NealEnsemble,  AffineEnsemble, Nishihara, Exchanging, kou2006, Liang00evolutionarymonte, MultiTry}), though as far as we are aware, the physical interpretation as well as the specific suite of algorithms
we propose is new. These methods fall generally into two categories: those that, like the suburban algorithm, operate over identical copies of the target distribution; and those that operate over a parameterized family of related but not identical distributions.  The former includes reference \cite{Geyer}, which uses a population of samples to adaptively choose a proposal direction; reference \cite{AffineEnsemble}, which use a population of samples to generate proposals that are invariant under affine transformations of the underlying space; and~\cite{Nishihara}, which uses subsets of the sample populations to estimate parameters of an approximate distribution used to generate proposals in an elliptical slice sampler.  The second category is exemplified by parallel tempering~\cite{ParallelTempering}, in which parallel chains operate over a family of distributions parameterized by temperature where proposals include both local transitions and exchanges of state between pairs of chains.  This category of methods uses distributions that mix better than the target distribution, but are similar enough to each other that exchanges will be accepted with reasonable probability.  There are many MCMC variations in the literature that follow this general approach, including references \cite{Exchanging, kou2006, Liang00evolutionarymonte, MultiTry}.

Returning to the case of suburban samplers, there are potentially many consistent ways to connect together the
inferences of statistical agents. From the perspective of physics, this
amounts to a notion of distance/proximity between nearest neighbors in a
brane. A physically well-motivated way to eliminate this arbitrary feature
is to allow the notion of proximity itself to \textit{fluctuate}. From the perspective of
equation (\ref{Qextend}), we treat the placement of nearest neighbors as
specifying a collection of random graphs, and by allowing possible
fluctuations, various agents reach a collective inference differently.
Indeed, from this perspective, it is also natural to allow the brane to split into
or join up smaller constituent parts (see figure \ref{parstringbrane}).
In contrast to the case of a grid with a fixed topology, the physics of general splitting and
joining is less tractable analytically (except in special limits where
perturbation theory via a small expansion parameter is available).

Turning the discussion around, the general considerations presented here appear to have consequences
for our understanding of quantum fields and strings. As noted in reference \cite{Heckman:2013kza}, one way to form
approximate observables in a theory of quantum gravity is to consider inference of an ensemble of
agents pooling their (approximate) local observations. From this perspective, the present paper can be viewed as a concrete
implementation of this general proposal using the framework of Markov Chain Monte Carlo sampling. In particular, the appearance
of a preferred role for an effective one-dimensional connectivity as predicted in \cite{Heckman:2013kza} suggests a
central role for such objects in any formulation of quantum gravity.

We view MCMC with extended strings and branes as a novel
class of ensemble samplers in which there is some random degree of
connectivity between multiple statistical agents. By correlating the
inferences of nearest neighbors in this way, we can expect there to be some impact
on performance. For example, the degree of
connectivity impacts the mixing rate for obtaining independent samples.
Another important feature is that because we are dealing with an
extended object, different statistical agents may become localized in
different high density regions. Provided the connectivity with neighbors is
sufficiently low, coupling these agents then has the potential to provide
a more accurate global characterization of a target distribution. Conversely,
connecting too many agents together may cause the entire collective to suffer
from \textquotedblleft groupthink\textquotedblright\ in the sense of
\cite{Heckman:2013kza}, namely, once an initial erroneous inference is reached it can
become difficult to correct. In the statistical mechanical interpretation of
statistical inference developed in references \cite{BalasubramanianGeo, Balasubramanian:1996bn, Heckman:2013kza},
this can be viewed as the standard tradeoff in thermodynamics between
minimizing the energy (i.e., obtaining an accurate inference) and
maximizing entropy (i.e., exploring a broader class of configuration).
In particular, we shall present some general arguments that the optimal
connectivity for a network of agents on a grid
arranged as a hypercubic lattice with some percolation (i.e., we allow for broken links)
occurs at a critical effective dimension:
\begin{equation}\label{crit}
d_{\mathrm{eff}} \sim 1
\end{equation}
where $2 d_{\mathrm{eff}} $ is the average number of neighbors.

To summarize: With too few friends one drifts into oblivion,
but with too many friends one becomes a boring conformist.

To test these general theoretical considerations, we perform a number of numerical
experiments for a variety of simple target distributions. One of the simple
features of this class of proposal kernels is that there is a
hyperparameter available (the average degree of connectivity) which allows us
to smoothly interpolate from the case of an extended object to a collection of
independent parallel MH samplers. Overall, we find that some level of connectivity leads to a generic
improvement over parallel MH.

We address the extent to which the extended nature of a brane impacts its performance.
Holding fixed the average effective dimension $d_{\mathrm{eff}}$
but varying the overall topology of the extended object from a 1d, to 2d, to
4d grid, as well as an Erd\"os-Renyi ensemble of random graphs leads to
comparable performance for the different samplers. In all of the cases we have
encountered, the mixing rate is indeed fastest at a critical effective dimension as dictated by line (\ref{crit}).
In some cases, however, the clumping effects of a higher
dimensional grid are helpful, especially when there is a landscape of local maxima
in the target distribution.

\begin{figure}[t!]%
\centering
\includegraphics[
scale = 0.5, trim = 0mm 0mm 0mm 0mm
]%
{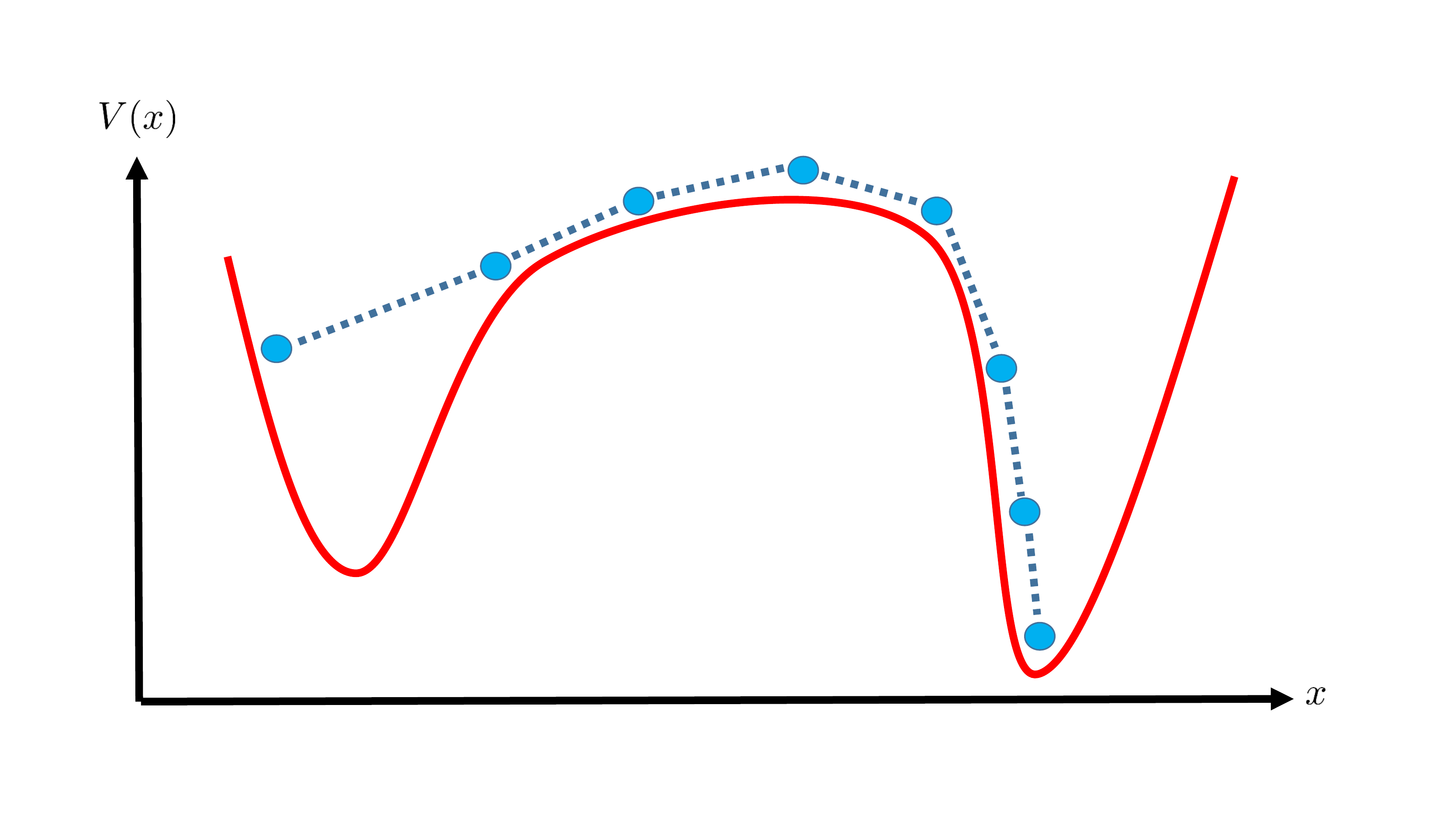}%
\caption{Depiction of how an extended object such as a string can overcome a free energy barrier.}%
\label{twodcubetaus}%
\end{figure}

The rest of this paper is organized as follows. We first begin in section
\ref{sec:STRING} with some general qualitative considerations
on the potential links between statistical inference and extended objects such as those which arise in string theory.
We then turn in section \ref{sec:MCEXTEND} with a general discussion on the physics of extended
objects and its relation to MCMC. Readers not interested in the theoretical underpinnings
of the algorithm can bypass most of section \ref{sec:MCEXTEND}. In section \ref{sec:SUBURBAN} we present
the \textquotedblleft suburban algorithm.\textquotedblright\  In
section \ref{sec:EXPERIMENT} we turn to an overview of our numerical experiments.
Section \ref{sec:DIMENSION} highlights the
dependence of the algorithm on the various hyperparameters, and in particular the average
degree of connectivity with neighbors. In sections \ref{sec:LANDSCAPE} and
\ref{sec:BANANA} we study particular examples of target distributions, and in
section \ref{sec:FREEBALL} we study some controlled examples where we increase
the free energy barrier between centers of a mixture model of two normal
distributions, showing that as the barrier separation increases, the
performance of parallel MH degrades more quickly than
a $d_\mathrm{eff} \sim 1$ suburban sampler. Section \ref{sec:CONC} contains our conclusions
and potential directions for future work. In Appendix \ref{app:SLICE}
we discuss in more detail the relative performance with slice sampling. For a condensed
account of our results, we refer the interested reader to reference \cite{Heckman:2016jud}.

Finally, a standalone copy of the \texttt{Java} libraries for the suburban sampler,
and its interface with the \texttt{Dimple} libraries is available
at the publicly available \texttt{GitLab} repository \texttt{https://gitlab.com/suburban/suburban}.
We have also included a short \texttt{Matlab} demo for the suburban sampler.

\section{Statistical Inference with Strings and Branes \label{sec:STRING}}

To frame the results to follow, in this section we discuss both the physical motivation and applications connected
with statistical inference with extended objects such as strings and branes.

The essential point is that in the context of a quantum theory of gravity such as string theory, it is
not entirely clear whether there is a completely well-defined notion of a local observable. Along these lines,
it is fruitful to ask whether a collective of observers can agree in some approximate way on data measured by an
ensemble. With this in mind, reference \cite{Heckman:2013kza} proposed to study the observations of a collective
of statistical agents pooling their resources to reach a final inference scheme. A concrete way to pose this question
is to ask the sense in which the collective can accurately reconstruct a joint probability distribution such as:
\begin{equation}
\pi(x_1,...,x_M) = \pi(x_1) ... \pi(x_M),
\end{equation}
for $M$ statistical agents. Following the general ideas presented in references \cite{Balasubramanian:1996bn, BalasubramanianGeo}
for individual agents, statistical inference can be understood in terms of a statistical mechanics problem in which the
relative entropy between an agents proposed probability distribution and the true distribution provides a notion of energy.
This suggests a natural application to quantum gravity, where an individual observer may only have access to their individual ``worldview''
which is then improved by further samples of an actual data set.

Now, in quantum gravity there is a well-known issue with the use of point particles which stems from the fact that there is strictly speaking, no
notion of a gauge invariant local observable. Rather, it is generally expected that some notion of locality must give way, and must also
be accompanied by the appearance of spread out or extended objects. Along these lines, it is natural to ask whether an inference scheme
adopted by an extended object can lead to different conclusions from those obtained by independent point particles.

This question was studied in reference \cite{Heckman:2013kza}, where general considerations led to the conclusion
that the standard conditions of quantum strings suggest a privileged role for one-dimensional objects.
The main idea of \cite{Heckman:2013kza} is that when statistical agents share data along a discretized worldvolume lattice, new inference schemes can be achieved which are unavailable to an individual agent. Additionally, there is a privileged role for $1+1$ dimensional objects
because in this case, the two-point function for a scalar field exhibits a late-time logarithmic divergence. This is milder than the power law divergence present for a free point particle, suggesting a more stable inference scheme relative to this case. Coupling this system to worldsheet gravity can also be understood at an abstract level as an additional layer of inference by
a meta-agent, namely, where the connectivity between nearest neighbors can be rearranged.

These general considerations naturally suggest a number of important followup questions, especially in the context of string theory. For
one, the privileged role of one-dimensional objects appears to be at odds with some of the general lessons reached from
the study of non-perturbative dualities, where various extended objects are in some sense on an ``equal footing'' with quantum strings. This in turn
raises the question of whether the connection between strings and inference in quantum gravity is only an artifact of working with a particular
geometric connectivity for agents in the collective. At a more concrete level, there is also the question of the precise mechanism by which a collective actually ``shares'' information, namely how the pooling of resources in the entire collective actually takes place.

One of the aims of the present paper will be to address these issues by showing how inference can actually be implemented
for an extended object. Along these lines, we focus on the case of Markov Chain Monte Carlo sampling methods. The appearance of worldvolume
gravity for the extended object will also be crudely characterized in terms of a statistical ensemble of random graphs, which act to define a time-dependent notion of locality for agents in the collective.
In a certain sense this is a cruder notion of gravity than is present in the physical superstring,
but it has the advantage of being discretized and fully non-perturbative. From this perspective, one should view the results of this paper
as a concrete way to implement a non-perturbative formulation of strings and branes making observations in a target space.

The average degree of connectivity will provide us with a notion of an effective dimension. While this is admittedly less refined than the standard
notions used in much of the high energy theory literature, it has the definite advantage of being completely well-defined so that we can implement and test it numerically. Indeed, even though it is crude, the remarkable fact that there is an effective dimension which appears to govern the main elements of the inference scheme is highly non-trivial and provides further evidence of the crucial role of effectively one-dimensional objects.

Finally, though we shall be implementing a Markov Chain Monte Carlo sampling algorithm the aim here is to better understand how the topology and dimension of a fluctuating lattice itself influences the overall speed and accuracy of an inference scheme. This is rather different from the
standard approach in lattice quantum field theory where it is typically assumed that the lattice is fixed, and moreover, the structure of the
target distribution $\pi(x)$ is assumed to take a relatively simple canonical form. With these physical considerations in mind, we now
turn to the implementation of MCMC with strings and branes.

\section{MCMC with Strings and Branes\label{sec:MCEXTEND}}

One of the main ideas we shall develop in this paper is MCMC methods for
extended objects. In this section we begin with the
theoretical elements of this proposal, giving a path integral formulation of MCMC for point particles and branes.
Some of this material is likely familiar to some physicists as the ``Feynman-Kac'' path integral formulation of
stochastic processes, though as far as we are aware, the specific application to MCMC methods we
focus on here has not appeared before in the literature. For earlier related work on
the statistical mechanics of statistical inference, see
\cite{BalasubramanianGeo, Balasubramanian:1996bn} and \cite{Heckman:2013kza}. For a relatively concise review of
some aspects of string theory and the physics of branes, we refer the interested reader to
\cite{Zwiebach:2004tj, Johnson:2005mqa}, and references therein. For additional background on details of quantum field
theory, we refer the interested reader to \cite{Peskin:1995ev, Wipf:2013vp}, and references therein.

Suppose then, that we have a target distribution $\pi(x)$. In an
MCMC\ algorithm we produce a sequence of \textquotedblleft
timesteps\textquotedblright\ $x^{(1)},...,x^{(t)},...,x^{(N)}$ which can be
viewed as the motion of a point particle exploring a target space $\Omega$. More
formally, this sequence of points defines the \textquotedblleft
worldline\textquotedblright\ for a particle, and consequently a map from time
to the target:%
\begin{equation}
x:\text{Worldline}\rightarrow\text{Target \ \ with \ \ }t\mapsto x(t).
\end{equation}
In the case of a string, we extend the notion of a \textquotedblleft
worldline\textquotedblright\ to a \textquotedblleft worldsheet,\textquotedblright\
i.e., we have both a temporal and spatial extent with respective coordinates $t$ and
$\sigma$:%
\begin{equation}
x:\text{Worldsheet}\rightarrow\text{Target \ \ with \ \ }(t,\sigma)\mapsto
x(t,\sigma)
\end{equation}
More generally, if we have an extended object with $d$ spatial directions,
we get a map from a \textquotedblleft worldvolume\textquotedblright\ to the
target:%
\begin{equation}
x:\text{Worldvolume}\rightarrow\text{Target \ \ with \ \ }(t,\sigma
_{1},...,\sigma_{d})\mapsto x(t,\sigma_{1},...,\sigma_{d}).
\end{equation}
The case $d=0$ and $d=1$ respectively denote a point particle and string. To
make the analysis of these maps computationally tractable, we will have to
discretize these worldvolumes. So, in addition to making finite timesteps, we
will also have to work with a finite number of statistical agents spanning the
spatial directions of the worldvolume.

Using this formulation, we shall extract some basic properties such as the correlation between
samples as a function of time. In particular, we will see that the
overall connectivity, i.e., the number of nearest neighbor interactions,
strongly influences both spatial as well as temporal correlations. This
spatial connectivity also affects the motion of the extended object on a fixed target.
Compared with the case of independent point particles, this can allow an extended
object to more easily explore global
aspects of a target.

The rest of this section is organized as follows. First, we give a path
integral formulation of MCMC for a point particle exploring
a fixed target distribution. We then turn to the generalization for strings
and branes, and introduce the notion of splitting and joining as well.
After introducing the general formalism, we then turn to an analysis of how
the average degree of connectivity for statistical agents in an ensemble
impacts the resulting inference scheme.

\subsection{Path Integral for Point Particles}

To frame the discussion to follow, in this subsection we introduce some
background formalism on path integrals. Our aim will be to gear up for the
case of extended objects.

In what follows, we denote the random variable as $X$ with outcome $x$ on a target space $\Omega$
with measure $dx$. We consider sampling from a probability density $\pi(x)$. In accord with physical
intuition, we view $- \log \pi(x)$ as a potential energy, i.e., we write:%
\begin{equation}
\pi(x)=\exp(-V(x)).
\end{equation}
In general, our aim is to discover the structure of $\pi(x)$ by using some
sampling algorithm to produce a sequence of values $x^{(1)},...,x^{(N)}$. A
quantity of interest is the expected value of $\pi(x)$ with respect to a given
probability distribution of paths. This helps in telling us the relative speed of convergence and the
mixing rate. To study this, it is helpful to evaluate the expectation value of
the quantity:%
\begin{equation}
\underset{i=1}{\overset{N}{%
{\displaystyle\prod}
}}\exp(-\beta^{(i)}V(x^{(i)}))
\end{equation}
with respect to a given path generated by our sampler. We can then
differentiate with respect to the $\beta^{(i)}$'s to study the rate at which
our sampler explores the target distribution.

In more general terms, the reason to be interested in this expectation value comes
from the statistical mechanical interpretation of statistical
inference \cite{BalasubramanianGeo, Balasubramanian:1996bn, Heckman:2013kza}: There is a natural
competition between staying in high likelihood regions (minimizing the potential), and exploring more of the distribution (maximizing
entropy). The tradeoff between the two is neatly captured by the path integral formalism. Indeed,
in the special case $\beta^{(i)} = 1$ we have an especially transparent interpretation: It tells us about
a particle moving in a potential $V(x)$, and subject to a thermal background, as specified by the
choice of probability measure over possible paths. Indeed, we will view this probability measure as defining a ``kinetic energy''
in the sense that each time step, we apply a random kick to the trajectory of the particle,
as dictated by its contact with the thermal reservoir.

Along these lines, if we have an MCMC\ sampler with transition
probabilities $T(x^{(i)}\rightarrow x^{(i+1)})$, the expected value depends on:
\begin{equation}
Z_{\text{path}}(\left\{  \beta^{(i)}\right\}  )=T(x^{(0)}\rightarrow
x^{(1)})e^{-\beta^{(1)}V(x^{(1)})}\times...\times T(x^{(N-1)}\rightarrow
x^{(N)})e^{-\beta^{(N)}V(x^{(N)})}%
\end{equation}
Marginalizing over the intermediate values, we get:
\begin{equation}
\mathcal{Z} = \int [dx] \left(
\underset{i=0}{\overset{N-1}{%
{\displaystyle\prod}
}}T(x^{(i)}\rightarrow x^{(i+1)})e^{- \beta^{(i+1)} V(x^{(i+1)})}\right)
\end{equation}
where we have introduced the measure factor $[dx] = dx^{(1)} ... dx^{(N)}$.
We would like to interpret $V(x)$ as a potential energy and $-\log
T(x^{(i)}\rightarrow x^{(i+1)})$ as a kinetic energy. So, we shall write:%
\begin{equation}
V(x)=-\log\pi(x)\text{ \ \ and \ \ }K(x^{(i)},x^{(i+1)})=-\log T(x^{(i)}%
\rightarrow x^{(i+1)}).
\end{equation}

We now observe that our expectation value has the form of a well-known object
in physics: A path integral!\footnote{Albeit one in Euclidean signature, see
below for details.} For example, with all $\beta^{(i)}=1$, we have:
\begin{equation}
\mathcal{Z}(x^{\text{begin}}\rightarrow x^{\text{end}})=\underset{\text{begin}%
}{\overset{\text{end}}{%
{\displaystyle\int}
}}[dx]\exp(- \underset{t}{\sum} L^{(E)}[x^{(t)}])
\end{equation}
where we have introduced the Euclidean signature Lagrangian:
\begin{equation}
L^{(E)}[x^{(t)}]=K+V.\text{ }%
\end{equation}
Since we shall also be taking the number of timesteps to be very large, we
make the Riemann sum approximation and introduce the rescaled Lagrangian
density:%
\begin{equation}
\frac{1}{N}\underset{t}{\sum}\mapsto\int dt,\text{ \ \ }NL^{(E)}%
\mapsto\mathcal{L}^{(E)}%
\end{equation}
so that we can write our process as:%
\begin{equation}
\mathcal{Z}(x^{\text{begin}}\rightarrow x^{\text{end}})=\int[dx]\exp\left(
-\int dt\mathcal{L}^{(E)}[x(t)]\right)  ,
\end{equation}
where by abuse of notation, we use the same variable $t$ to reference both the
discretized timestep as well as its continuum counterpart.

A few comments are in order here. Readers familiar with the Lagrangian
formulation of classical mechanics and quantum mechanics will note that we
have introduced $K+V$ rather than $K-V$ as our Lagrangian. In physical terms,
this has important consequences, particularly in the interpretation of the
time evolution of a saddle point solution (i.e., one that is solved by the
Euler-Lagrange equations of motion). As an illustrative example, we see that
for a quadratic potential, we do not obtain the familiar behavior of a
harmonic oscillator with trajectory $x(t)\sim\exp(i\omega t)$, but rather
$x(t)\sim\exp(-\omega t)$. Formally, this amounts to the substitution
$t\mapsto it$, which is often referred to as a \textquotedblleft Wick
rotation\textquotedblright\ or passing to \textquotedblleft Euclidean
signature.\textquotedblright\ Physically, what it means is that rather than
getting oscillatory behavior, we instead get a diffusion or spread to the
location of the particle. For further discussion on Euclidean signature
quantum field theory, i.e., statistical field theory, see for example
\cite{Peskin:1995ev, Wipf:2013vp}.

To give further justification for this terminology, consider now the specific
case of the Metropolis-Hastings algorithm. In this case, we have a proposal
kernel $q(x^{\prime}|x)$, and acceptance probability:%
\begin{equation}
a(x^{\prime}|x)=\min\left(  1,\frac{q(x|x^{\prime})}{q(x^{\prime}|x)}\frac
{\pi(x^{\prime})}{\pi(x)}\right)  .
\end{equation}
The total transmission probability is then given by a sum of two terms. One is
given by $a(x^{\prime}|x)q(x^{\prime}|x)$, i.e., we accept the new sample. We
also sometimes reject the sample, i.e., we keep the same value as before:%
\begin{equation}
T(x\rightarrow x^{\prime})=r\times\delta(x-x^{\prime})+a(x^{\prime
}|x)q(x^{\prime}|x),
\end{equation}
where $\delta(x - x^{\prime})$ is the Dirac delta function, and we have introduced
an averaged rejection rate:
\begin{equation}
r \equiv1-\int dx^{\prime}\text{ }a(x^{\prime}|x)q(x^{\prime}|x).
\end{equation}
The specific optimal value depends on the target distribution and the proposal kernel.\footnote{For example, under the
assumption that the limiting diffusion approximation is valid, the optimal acceptance rate is $0.234$ \cite{Gelman:1997}.}

For illustrative purposes, suppose that we work in the special limit where the acceptance rate is
close to one, and that we have a Gaussian proposal kernel so that $-\log
q\left(  x^{(t+1)}|x^{(t)}\right)  \sim\alpha\left(  x^{(t+1)}-x^{(t)}\right)
^{2}$. In this case, the path integral takes a rather pleasing form which has
a simple physical interpretation. We have:
\begin{equation}
\text{High Acceptance: }L^{(E)}[x^{(t)}]=K+V\simeq\alpha\left(  x^{(t+1)}-x^{(t)}\right)
^{2}+V(x^{(t)}).
\end{equation}
Where we interpret the finite difference between time steps as a time derivative:
\begin{equation}
D_{t}x\equiv x^{(t+1)}-x^{(t)}.
\end{equation}

More generally, we can ask what happens for intermediate values of $a$. In
general, this is a challenging question so we do not expect to have as simple
a form for the Euclidean signature Lagrangian.\ Nevertheless, we shall see
that much of the structure already found persists in this case as well. Along
these lines, we shall attempt to approximate the mixture model $T(x\rightarrow
x^{\prime})$ by a normal distribution $q_{\text{eff}}\left(  x^{(t+1)}%
|x^{(t)}\right)  $ such that  $-\log q_{\text{eff}}\left(  x^{(t+1)}%
|x^{(t)}\right)  \sim\alpha_{\text{eff}}\left(  x^{(t+1)}-x^{(t)}\right)
^{2}$. To this end, we match the first and second moments of the
putative normal distribution with our net transition rate in the approximation that
we can use the average acceptance rate $\overline{a}$:
\begin{equation}
\alpha_{\text{eff}}=\frac{1}{\overline{a}}\times\alpha.
\end{equation}
So in this more general case, we get the effective Lagrangian:%
\begin{equation}\label{generalLag}
L^{(E)}[x^{(t)}]\simeq\alpha_{\text{eff}}\left(  x^{(t+1)}-x^{(t)}\right)
^{2}+V(x^{(t)})+...,
\end{equation}
where here, the \textquotedblleft...\textquotedblright\ denotes additional
correction terms coming from:
\begin{equation}\label{TheCorrections}
\text{Correction Term: } - \log(T(x\rightarrow x^{\prime})/q_{\text{eff}%
}\left(  x^{(t+1)}|x^{(t)}\right)  ).
\end{equation}
At a large number of samples, we expect that contributions given by higher order powers of time derivatives
are suppressed by powers of $1/N$, a point we discuss in more detail in subsection \ref{ssec:SCALING}.
Observe that as the acceptance rate decreases $\alpha_{\text{eff}}$ increases and the sampled
values all concentrate together.

Our plan in the following sections will be to assume the structure of
a kinetic term with quadratic time derivatives, but a general
potential. The overall strength of the kinetic term will depend
on details such as the average acceptance rate. As we discuss in subsection
\ref{ssec:SCALING}, the correction terms to this general structure will, for a broad class of models,
be suppressed by powers of $1/N$.

\subsection{Path Integral for Extended Objects}

We now turn to the generalization of the above concepts for strings and
branes, i.e., extended objects. To cover this more general class of
possibilities, we first introduce $M$ copies of the original distribution, and
consider the related joint distribution:%
\begin{equation}
\pi(x_{1},...,x_{M})=\pi(x_{1})...\pi(x_{M})\text{.}%
\end{equation}
If we keep the proposal kernel unchanged, we can simply describe the evolution
of $M$ independent point particles exploring an enlarged target space:%
\begin{equation}
\Omega_{\text{enlarged}}= \Omega^{M}=\underset{M}{\underbrace{\Omega \times...\times \Omega}
}\text{.}%
\end{equation}
If we also view the individual statistical agents on the worldvolume as
indistinguishable, we can also consider quotienting by the symmetric group on
$M$ letters, $S_{M}$:%
\begin{equation}
\Omega^{\mathcal{S}}_{\text{enlarged}}=X^{M}/S_{M}\text{.}%
\end{equation}

Of course, we are also free to consider a more general proposal kernel in
which we correlate these values. Viewed in this way, an extended object is a
single point particle, but on an enlarged target space. The precise way in
which we correlate entries across a grid will in turn dictate the type of
extended object.

We begin with some general definitions, and later specialize to more
tractable cases. Along these lines, suppose we have a graph consisting
of $M$ nodes, and a corresponding undirected adjacency matrix $A$, which
consists of ones on the diagonal, and just zeroes and ones off the diagonal.
We also use $\sigma$ as a general index holder for \textquotedblleft spatial position\textquotedblright%
\ on a grid. We say that two nodes $\sigma$ and $\sigma^{\prime}$ are \textquotedblleft
neighbors\textquotedblright\ if $A_{\sigma,\sigma^{\prime}}=1$. Denote by
Nb$(\sigma)$ the set of neighbors for site $\sigma$. Holding fixed the adjacency matrix $A$, we define the proposal
kernel:%
\begin{equation}
q(x_{1},...,x_{M}|y_{1},...,y_{M},A) \equiv \underset{\sigma=1}{\overset{M}{%
{\displaystyle\prod}
}}q_{\sigma}(x_{\sigma}|y_{\text{Nb}(\sigma)}),
\end{equation}
where $q_{\sigma}$ is some choice of proposal kernel for a single point particle.

We can therefore adopt two different perspectives on this procedure. On the
one hand, we can view an extended object as propagating over the enlarged
target. On the other hand, we can view this extended object as one collective
moving on the original target space.

Indeed, much of the path integral formalism carries over unchanged. The only
difference is that now, we must also keep track of the spatial extent of our
object. So, we again introduce a potential energy $V$ and a kinetic energy $K$:
\begin{equation}
V = -\log\pi \text{ \ \ and \ \ }K = -\log T,
\end{equation}
and a Euclidean signature Lagrangian density:
\begin{equation}
L^{(E)}[x(t,\sigma_{A})]=K+V,
\end{equation}
where here, $\sigma_{A}$ indexes locations on the extended object, and the
subscript $A$ makes implicit reference to the adjacency on the graph. The
transition probability is:
\begin{equation}
\mathcal{Z}(x_{\text{begin}}\rightarrow x_{\text{end}} | A)=%
{\displaystyle\int}
[dx]\text{ }\exp(-\underset{t}{\sum}\underset{\sigma}{\sum
}L^{(E)}[x(t,\sigma_{A})]),
\end{equation}
where now the measure factor $[dx]$ involves a product over $dx^{(t)}_{\sigma}$.
Since we shall also be taking the number of time steps and agents to be large, we
again make the Riemann sum approximation and introduce the rescaled Lagrangian
density:%
\begin{equation}
\frac{1}{N}\underset{t}{\sum}\mapsto\int dt,\text{ \ \ }\frac{1}%
{M}\underset{\sigma}{\sum}\mapsto\int d\sigma_{A}\text{ \ \ }NML^{(E)}%
\mapsto\mathcal{L}^{(E)}
\end{equation}
so that the expectation value has continuum description:
\begin{equation}
\mathcal{Z}(x^{\text{begin}}\rightarrow x^{\text{end}} | A)=\int%
[dx]\exp\left(  -\int dtd\sigma_{A}\text{ }\mathcal{L}%
^{(E)}[x(t,\sigma_{A})]\right)  ,
\end{equation}
in the obvious notation. Strictly speaking, the integral with measure $d\sigma_{A}$ may fail to have a
smooth continuum limit (i.e., when $M\rightarrow\infty$), so when it does not,
no continuum approximation is available and we should view this as merely a
shorthand for the discretized answer.

\subsubsection{Splitting and Joining \label{ssec:wgravity}}

In the above discussion, we held fixed a particular choice of adjacency
matrix. This choice is somewhat arbitrary, and physical considerations suggest
a natural generalization where we sum over a statistical ensemble of choices.
We shall loosely refer to this splitting and joining of connectivity
as \textquotedblleft incorporating gravity\textquotedblright\ into the dynamics of the extended object, because
it can change the notion of which statistical agents are nearest
neighbors.\footnote{It is not quite gravity in the worldvolume theory, because
there is a priori no guarantee that our sum over different graph topologies
will have a smooth semi-classical limit. Nevertheless, summing over different
ways to connect the statistical agents conveys the main point that the
proximity of any two agents can change. For additional discussion,
see for example reference \cite{Heckman:2013kza}.} Along these lines, we incorporate an
ensemble $\mathcal{A}$ of possible adjacency matrices, with some prescribed
probability to draw a given adjacency matrix. Since we evolve forward in
discretized time steps, we can in principle have a sequence of such matrices
$A^{(1)},...,A^{(N)}$, one for each timestep. For each draw of an adjacency
matrix, the notion of nearest neighbor will change, which we denote by writing
$\sigma_{A(t)}$, that is, we make implicit reference to the connectivity of
nearest neighbors. Marginalizing over the choice of adjacency matrix, we get:%
\begin{equation}
\mathcal{Z}(x_{\text{begin}}\rightarrow x_{\text{end}})=%
{\displaystyle\int}
[dx][dA]\text{ }\exp(-\underset{t}{\sum}\underset{\sigma
}{\sum}L^{(E)}[x(t,\sigma_{A(t)})]),
\end{equation}
where now the integral involves summing over multiple ensembles:
the spatial and temporal values with measure factor $dx_{\sigma}^{(t)}$, as well as the choice
of a random matrix from the ensemble with measure factor $dA^{(t)}$ (one such integral for each
timestep). At a very general level, one can view the adjacency matrix as
adding additional auxiliary random variables to the process. So in this sense,
it is simply part of the definition of the proposal kernel.

The topology of an extended object dictates a
choice of statistical ensemble $\mathcal{A}$. We illustrate this by giving some
particular examples which we study in more detail later on. For a collection
of $M$ independent, but indistinguishable point particles, the ensemble of
adjacency matrices is given by:%
\begin{equation}
\mathcal{A}_{\text{particles}}=\left\{  SAS^{-1}|A\text{ is the }M\times
M\text{ identity and }S\in S_{M}\right\}  .
\end{equation}
For an ensemble of strings, we have a notion of a nearest neighbor
interaction, and so we also introduce a split / join probability
$p_{\text{join}}$:%
\begin{equation}
\mathcal{A}_{\text{string}}(p_{\text{join}})=\left\{  SAS^{-1}\text{ with:\ }%
\begin{array}
[c]{l}%
S\in S_{M}\text{,}\\
A_{\sigma\sigma}=1\text{,}\\
A_{\sigma,\sigma+1}=A_{\sigma+1,\sigma}=1\text{ with probability }p_{\text{join}}\text{,}\\
A_{\sigma\sigma^{\prime}}=0\text{ otherwise}%
\end{array}
\right\}  ,
\end{equation}
where in the above, the index $\sigma=M+1$ is identified with $\sigma=1$. That
is, we have a circulant matrix: Geometrically, we view $1,...,M$ as arranged
along a circle, with each link either on or off.

More generally, we can consider the case of a $d$-dimensional hypercubic lattice,
i.e., an extended object in $d$ spatial dimensions. In this case, it is
somewhat simpler to first introduce a $\underset{d}{\underbrace{m\times
...\times m}} \times \underset{d}{\underbrace{m\times
...\times m}}$ array with $m^{d}=M$, which we then repackage in terms of an
$M\times M$ matrix. For a hypercubic lattice in $d$ dimensions, we introduce
$A_{\sigma_{1},...,\sigma_{d};\sigma_{1}^{\prime},....,\sigma_{d}^{\prime}}$,
and define the ensemble of arrays for a brane as:%
\begin{equation}
\mathcal{A}_{\text{brane}}(p_{\text{join}})=\left\{  S A S^{-1}\text{ with:\ }%
\begin{array}
[c]{l}%
S\in S_{M}\text{,}\\
A_{\sigma_{1},...,\sigma_{d};\sigma_{1},...,\sigma_{d}}=1\text{,}\\
A_{\sigma_{1},...,\sigma_{k},...\sigma_{d};\sigma_{1},...,\sigma_{k}%
+1,...,\sigma_{d}}=\\
A_{\sigma_{1},...,\sigma_{k}+1,...\sigma_{d};\sigma_{1},...,\sigma_{k}
,...,\sigma_{d}}=1\text{ with probability }p_{\text{join}}\text{,}\\
A_{\sigma\sigma^{\prime}}=0\text{ otherwise}%
\end{array}
\right\}  .
\end{equation}
We can repackage this as an $M\times M$ adjacency matrix by replacing the
multi-index $\sigma_{1},...,\sigma_{d}$ by a single base $m$ index:%
\begin{equation}
i=1+(\sigma_{1}-1)+(\sigma_{2}-1)m+...+(\sigma_{d}-1)m^{d-1}.
\end{equation}

Of course, in addition to these geometrically well-motivated choices, we can
consider more general ensembles of adjacency matrices. For example, a
configuration of random graphs with well studied properties is the
Erd\"{o}s-Renyi ensemble:%
\begin{equation}
\mathcal{A}_{\text{ER}}(p_{\text{join}})=\left\{
\begin{array}
[c]{l}%
A_{\sigma\sigma}=1\text{,}\\
A_{\sigma\sigma^{\prime}}=A_{\sigma^{\prime}\sigma}= 1\text{ with probability }p_{\text{join}}\text{
(}\sigma\neq\sigma^{\prime}\text{)}%
\end{array}
\right\}  .
\end{equation}

\subsection{Dimensions and Correlations} \label{ssec:SCALING}

In the previous section we presented some general features of strings and
branes, and their generalization to Markov chains. Following some of the
general considerations outlined in reference \cite{Heckman:2013kza}, in this
section we discuss the extent to which the extended nature of such objects
plays a role in statistical inference and in particular MCMC.

To keep our discussion from becoming overly general, we shall initially
specialize to the case of a hypercubic lattice of agents in $d$ spatial dimensions
arranged on a torus, and we denote a location on the grid by a $d$-component
vector $\sigma$. We shall later relax these considerations to
allow for the possibility of a fluctuating worldvolume.

For a fixed grid, each grid site has precisely $2d$ neighbors. In what
follows, we shall find it convenient to introduce a set of $d$ unit vectors:%
\begin{align}
e_{1} &  =(1,0,...,0)\\
e_{2} &  =(0,1,...,0)\\
&  ...\\
e_{d} &  =(0,0,...,1).
\end{align}
We also specialize the form of the proposal kernel:
\begin{equation}
q_{\sigma}(x_{\sigma}(t+1)|\text{Nb}%
(x_{\sigma}(t)))\propto \exp\left(
\begin{array}
[c]{l}%
-\alpha\left(  x_{\sigma}(t+1)-x_{\sigma
}(t)\right)  ^{2}\\
-\underset{k=1}{\overset{d}{\sum}}\beta\left(  x_{\sigma
}(t+1)-x_{\sigma+e_{k}}(t)\right)  ^{2}\\
-\underset{k=1}{\overset{d}{\sum}}\beta\left(  x_{\sigma
}(t+1)-x_{\sigma-e_{k}}(t)\right)  ^{2}%
\end{array}
\right)  .
\end{equation}
This has a recognizable form, consisting of finite differences in both the time
direction, and spatial directions of our brane. Along these lines, we
introduce the notation:%
\begin{align}
D_{t}x_{\sigma} &  =x_{\sigma}
(t+1)-x_{\sigma}(t)\\
D_{+k}x_{\sigma} &  =x_{\sigma + e_{k}}(t)-x_{\sigma}(t)\\
D_{-k}x_{ \sigma} &  =x_{\sigma - e_{k}}(t)-x_{\sigma}(t)
\end{align}
so that the proposal kernel is given by:%
\begin{equation}
q_{\sigma}(x_{\sigma}|\text{Nb}%
(x_{\sigma}))\propto \exp\left(  -\alpha\left(  D_{t}%
x_{\sigma}\right)  ^{2}-\underset{k=1}{\overset{d}{\sum}%
}\beta\left(  D_{t}x_{\sigma}-D_{+k}x_{\sigma}\right)  ^{2}-\underset{k=1}{\overset{d}{\sum}}\beta\left(  D_{t}%
x_{\sigma}-D_{-k}x_{\sigma}\right)
^{2}\right)  .
\end{equation}
To proceed further, we observe that in a large lattice, the finite differences
are well-approximated by derivatives of continuous functions. In this case, we
can also write $D_{+k}x_{\sigma}=-D_{-k}x_{\sigma}$, up to higher order
derivatives, which as we explain in subsection \ref{ssec:SCALING} make a subleading
contribution to the inference problem. Expanding in this limit, various
cross-terms cancel and we get:
\begin{equation}
q_{\sigma}(x_{\sigma}|\text{Nb}%
(x_{\sigma}))\propto\exp\left(  -(\alpha+2d\beta)\left(
D_{t}x_{\sigma}\right)  ^{2}-\underset{k=1}{\overset{d}{\sum
}}2\beta\left(  D_{+k}x_{\sigma}\right)  ^{2}\right)  .
\end{equation}
That is, we see the expected kinetic term for a $(d+1)$-dimensional quantum
field theory in Euclidean signature.

So far, we have kept our analysis rather general. Now, we would also like to be able
to take a canonical limit in which the strength of timelike jumps remains comparable in passing from the
completely disconnected grid to the maximally connected grid. To this end, we now further specialize
the choice of $\alpha$ as:
\begin{equation}
\alpha=2\beta-2d\beta.
\end{equation}
The full proposal kernel now takes the form:
\begin{equation}
\underset{\sigma}{%
{\displaystyle\prod}
}q_{\sigma}(x_{\sigma}|\text{Nb}%
(x_{\sigma}))\propto\exp\left(  -2\beta
\underset{\sigma}{%
{\displaystyle\sum}
}\left(  \left(  D_{t}x_{\sigma}\right)  ^{2}%
+\underset{k=1}{\overset{d}{\sum}}\left(  D_{+k}x_{\sigma
}\right)  ^{2}\right)  \right)  .
\end{equation}

Now, just as in the case of the point particle path integral, we again see that the effective transition rate defines
a kinetic energy term, with an effective strength dictated by the overall acceptance rate. The general
form of this kinetic term is given by a form recognizable to physicists:\footnote{As the astute reader will
no doubt notice, the structure of the kinetic term we consider here is not the most general one
we could consider. More generally, we can introduce a vector of temporal and spatial derivatives $D_{K} x$, with $K = 0,1,...,d$
and introduce the kinetic term $\frac{1}{2}(D_{K} x) \left(\frac{1}{\Sigma} \right)^{K L} (D_{L} x) $, with $\Sigma$ a positive definite matrix. In the physics literature, this defines a metric on the brane system. An even further generalization is to allow some $x$ dependence in $\Sigma$ itself. Some aspects of this more general case were considered in \cite{Heckman:2013kza}. We leave a detailed study
of the application to MCMC for future work.}
\begin{equation}
L^{(E)}[x(t,\sigma)]=2\beta_{\mathrm{eff}}\underset{\sigma}{%
{\displaystyle\sum}
}\left(  \left(  D_{t}x_{\sigma}\right)  ^{2}%
+\underset{k=1}{\overset{d}{\sum}}\left(  D_{+k}x_{\sigma
}\right)  ^{2}\right)  +V + ...
\end{equation}
where $\beta_{\mathrm{eff}}$ sets the effective tension of the brane, and the correction terms ``...'' indicate that
we are again working to quadratic order in the derivatives. So to summarize, we have arrived at a
($d+1$)-dimensional statistical field theory with kinetic term quadratic in derivatives and a general potential.

One of the things we would most like to understand is the extent to which an
extended object with $d$ spatial dimensions can explore the hills and valleys
of $V$. We perform a perturbative analysis, at first viewing $V$ as a small correction to the
propagation of our extended object. Starting from some fixed position
$x_{\ast}$, we can then consider the expansion of $V$ around this point:%
\begin{equation}
V(x)=V(x_{\ast})+ V^{\prime}(x_{\ast}) (x - x_{\ast}) + \frac{V^{\prime\prime}(x_{\ast})}{2} (x - x_{\ast})^2 + ...,
\label{PotentialExpand}%
\end{equation}
and study the impact on the correlation of samples as a function of time. Each
of the derivatives of $V(x)$ reveals another characteristic feature length of
$V(x)$. These feature lengths are specified by the values of the moments
for the distribution $\pi(x)$. Alternatively, we can simply use the various derivatives of $V(x)$ to extract this set:
\begin{equation}
\ell_{n} \sim \frac{1}{\vert V^{(n)}(x_{\ast}) \vert ^{1/n}},
\end{equation}
An infinite value for the feature length simply means there is no new feature length.

Let us refer to the set of finite characteristic length scales as $\{ \ell_{i} \}$. Now, there is a clear sense in
which we can also view each of these length scales as defining a unit of time on the brane, i.e., how fast we expect our
sampler to explore such a feature length. Using our Lagrangian interpretation,
these length scales are set by both $\ell_{i}$ and the strength of the kinetic term:
\begin{equation}
\tau_{i} \sim \ell_{i} \times \sqrt{\beta}.
\end{equation}
We refer to ``early'' and ``late'' time behavior as specified by:
\begin{equation}
t_{\text{early}} \ll \tau_{i} \ll t_{\text{late}}.
\end{equation}
Since space and time on the worldvolume are on a similar footing, this also defines a notion of
``close'' and ``far'' for agents on the grid. By abuse of terminology, we shall lump all of these notions together.

Now, in the limit where $V=0$, there is a well-known behavior for correlation functions:
\begin{equation}
\left\langle x(t,\sigma)x(0,0)\right\rangle \equiv \frac{1}{\mathcal{Z}}%
{\displaystyle\int}
[dx]\text{ }x(t,\sigma)x(0,0)\text{ }\exp
(-\underset{t}{\sum}\underset{\sigma}{\sum}L^{(E)}[x(t,\sigma_{A})])
\label{corrfunc}%
\end{equation}
which for $(t, \sigma) \in \mathbb{R}^{d+1}$ is given by:\footnote{One way to obtain this scaling relation is to observe that the Fourier transform
of $1/k^2$ in $d+1$ dimensions exhibits the requisite power law behavior.}
\begin{equation}
\left\langle x(t,\sigma)x(0,0)\right\rangle \sim\frac{1}{\left(\sqrt{
t^{2}+\underset{i=1}{\overset{d}{\sum}}\left(  \sigma_{i}\right)  ^{2} } \right)
^{d-1}}.
\end{equation}
There is thus a rather sharp change in the behavior of the extended object for
$d<1$ and $d>1$. For $d=1$, we have a logarithm rather than a constant. So,
for low enough values of $d$, the extended object can wander around at late times, while for
larger values, the overall spread in values is suppressed.\ The crossover
between the two behaviors occurs at $d = 1$, i.e., the case of a string.

We would now like to understand the impact that adding a non-trivial potential energy will have on the structure of
our correlation functions. In general, this is a challenging problem which has no closed form solution. We can, however,
develop a picture for whether we expect these perturbations to impact the early and late time behavior of our sampler. Along these lines,
we can introduce the notion of a ``scaling dimension'' for $x(t,\sigma)$ and its derivatives. The basic idea is that just as we assign a notion of
proximity in space and time to agents on a grid, we can also ask how rescaling all distances on the grid via:
\begin{equation}\label{rescaler}
N \mapsto \lambda N \,\,\, M \mapsto \lambda^{d} M
\end{equation}
impacts the structure of our continuum theory Lagrangian.
The key point is that provided $N$ and $M$ have been taken sufficiently
large, or alternatively we take $\lambda$ sufficiently large, we
do not expect there to be any impact on the physical interpretation.

Unpacking this statement naturally leads us to the notion of a
scaling dimension for $x(t, \sigma)$ itself. Observe that
rescaling the number of samples and number of agents in
line (\ref{rescaler}) can be interpreted equivalently as holding
fixed $N$ and $M$, but rescaling $t$ and $\sigma$:
\begin{equation}
(t,\sigma) \mapsto (\lambda t , \lambda \sigma).
\end{equation}
Now, for our kinetic term to remain invariant, we need to \textit{also} rescale $x(t,\sigma)$:
\begin{equation}
x(t, \sigma) \mapsto \lambda^{- \Delta} x( \lambda t , \lambda \sigma).
\end{equation}
The exponent $\Delta$ is often referred to as the ``scaling dimension'' for $x$ obtained from ``naive dimensional analysis'' or NDA.
It is ``naive'' in the sense that when the potential $V \neq 0$ and we have strong coupling, the notion of a scaling dimension may only emerge at sufficiently long distance scales. For additional discussion on scaling dimensions and their role in statistical field theory, we refer the
interested reader to reference \cite{DiFrancesco:1997nk}. Note that because we
are uniformly rescaling the spatial and temporal pieces of the grid, we get the same answer for the scaling dimension if we consider spatial derivatives along the grid. This assumption can also be relaxed in more general physical systems.

To illustrate, let us now extract the scaling dimension of $x$ for the case
of a kinetic term quadratic in derivatives. We take $Dx$ as a placeholder for any choice of
derivative either in space or in time. Under a rescaling, we have:
\begin{equation}
\int dt d^{d} \sigma \text{ } (D x)^2 \mapsto  \lambda^{- 2 \Delta + d-1} \int dt d^{d} \sigma \text{ } (D x)^2.
\end{equation}
So, invariance of the action requires the exponent of $\lambda$ to vanish, namely:
\begin{equation}\label{freedim}
\Delta = \frac{d - 1}{2}.
\end{equation}

Using this general sort of scaling analysis allows us to characterize
possible effects of perturbations, and whether we expect them to drastically
impact our inference scheme as we take $N$ and $M$ to be very large.
As a first example, consider the effects of the ``Correction Terms'' in line
(\ref{TheCorrections}). We expect that such contributions will take the form of higher powers in $D x$, possibly multiplied by
powers of $x$ as well. The latter possibility mainly occurs when we have a proposal kernel which cannot be
written as temporal and spatial derivatives on a grid, i.e., it plays less of a role in the considerations that follow.

So, with this mind, we can consider the behavior of a perturbation of the form $(x)^{\mu} (D x)^{\nu}$.
Applying our NDA analysis prescription, we see that under a rescaling, the contribution such a term makes to the action is:
\begin{equation}
\int dt d^{d} \sigma \text{ } (x)^{\mu} (D x)^\nu \mapsto  \lambda^{-\mu \Delta - \nu (\Delta + 1) + d + 1} \int dt d^{d} \sigma \text{ } (D x)^2,
\end{equation}
However, using (\ref{freedim}), we see that the overall exponent on the righthand side is:
\begin{equation}
-\mu \Delta - \nu (\Delta + 1) + d + 1 = \frac{(2 - \nu) (d + 1) -\mu (d - 1)}{2}.
\end{equation}
So in other words, terms of the form $(D x)^{\nu}$ for $\nu > 2$ die off
as we take $N \rightarrow \infty$, i.e., $\lambda \rightarrow \infty$. Additionally, we
see that when $d \leq 1$, we can in principle expect more general contributions of
the form $(x)^{\mu} (D x )^{\nu}$. The presence of such terms will not affect our general
conclusions. For additional discussion on the interpretation of such contributions,
see reference \cite{Heckman:2013kza}.

Consider next possible perturbations to the potential energy.
Again, the impact these higher order terms can have
on the early time behavior of correlation functions of line (\ref{corrfunc})
depends on the number of dimensions for the brane. The main point follows from NDA:
In general, we are integrating over a $(d+1)$-dimensional spacetime,
so since a derivative carries one unit of inverse
length, the scaling dimension of $x$ (around the $V=0$ limit) is just
$(d-1)/2$. Each successive interaction term in the potential is of the form
$x^{n}$, with scaling dimension $n(d-1)/2$. As follows from a perturbative
analysis, when these higher order terms have low scaling dimension, their
impact on long distance correlations is strong, while conversely, when their scaling
dimension is high, their impact on long distance correlations is small.
The dividing line is set by whether the scaling dimension of the
interaction term is smaller than $d+1$ (i.e., the number of spacetime
directions we integrate over):%
\begin{equation}
\frac{n(d-1)}{2}\leq d+1.
\end{equation}
So, for $d\leq1$, all higher order terms can impact the long distance behavior
of the correlation functions, while for $d>1$, the most relevant term is
bounded above by:%
\begin{equation}
n\leq \frac{2d+2}{d-1}.
\end{equation}

Now, in the context of MCMC, we would like for our extended object to be able
to explore different contours of the energy landscape. This in turn means that
if our brane has settled near a critical point, it is potentially sensitive to
the higher order derivatives in $V(x)$ as in equation (\ref{PotentialExpand}).
So, a priori, if $V(x)$ possesses many non-trivial derivatives, taking $d\leq1$
provides a way to explore more of this landscape. More precisely, we can see
that for sufficiently large $d$ we cannot probe much of the global structure of
the potential. For example, if we set $n=3$, we see that $d\leq5$, i.e., six
spacetime dimensions for the worldvolume.

On the other hand, there is also a strong argument to avoid taking $d$ too
small. The fact that the time dependence of the two-point function of a free Gaussian
field goes as $1/t^{d-1}$ means that there can be significant spread in the
fluctuations of a low-dimensional object. This in turn means that such an
object may execute a very long random walk before finding anything of interest
(wandering in the desert).

So to summarize, for $d$ sufficiently small (i.e., close to zero), we can
expect to wander for a long time before finding anything of interest, while
conversely, if $d$ is bigger than one,
\textquotedblleft groupthink\textquotedblright\ takes over in the
collective and it is impossible to move away from an initial inference.

Clearly, the value of $d$ which is optimal will depend on the
precise shape of the potential $V(x)$. Nevertheless, we can already see that
there is potentially a significant advantage to correlating the behavior of
nearest neighbor interactions.

\subsubsection{Effective Dimension and Fluctuating Worldvolumes}

In the preceding discussion, we assumed that we had a fixed spatial grid of
dimension $d$, where the number of nearest neighbor interactions is always
fixed. There are a few drawbacks to this from the perspective of inference.
For example an extended object may become trapped more easily if all of its agents
clump in one local minimum of $V(x)$. On the
other hand, one of the advantages of an extended object is that there is a
natural pull to nearby minima, so it can also potentially explore a landscape
more efficiently than parallel point particles. To address this issue,
we consider a fluctuating worldvolume, i.e., we take an
ensemble of nearest neighbors which actually fluctuates as a function of time.

Since we are now dealing with a fluctuating number of nearest neighbors, we
will need to modify our proposal kernel. We again introduce a set of finite differences, but
now we specifically indicate the neighbor as $n(\sigma)$:
\begin{align}
D_{t}x_{\sigma} &  =x_{\sigma}(t+1)-x_{\sigma}(t)\\
D_{n(\sigma)}x_{\sigma} &  = x_{n(\sigma)}(t) - x_{\sigma}(t),
\end{align}
in the obvious notation. We now introduce a modified proposal kernel where the size of the
time step $\alpha_{\sigma}$ now depends on the number of neighbors:
\begin{align}
q_{\sigma}(x_{\sigma}|\text{Nb}(x_{\sigma})) &  \propto \exp\left(  -\alpha_{\sigma} \left(
D_{t}x_{\sigma}\right)  ^{2}-\underset{n(\sigma)}{\sum}\beta\left(  D_{t}x_{\sigma
} - D_{n(\sigma)}x_{\sigma}\right)  ^{2}\right)  \\
&  =\exp\left(  -(\alpha_{\sigma} + n_{\sigma}^{\mathrm{tot}} \beta)\left(  D_{t}x_{\sigma}\right)  ^{2}%
-\underset{n(\sigma)}{\sum}\beta\left(  D_{n(\sigma)}x_{\sigma}\right)
^{2} + \underset{n(\sigma)}{\sum} 2 \beta D_{t}x_{\sigma}D_{n(\sigma)}x_{\sigma
}\right)  ,
\end{align}
where $n_{\sigma}^{\mathrm{tot}}$ denotes the total number of nearest neighbors to the site $\sigma$, and
the parameter $\alpha_{\sigma}$ also depends on the total number of nearest neighbors:
\begin{equation}
\alpha_{\sigma} = 2\beta-n_{\sigma}^{\mathrm{tot}}\beta.
\end{equation}

Due to the fluctuating topology, an analysis of the correlation functions is now more challenging.
However, there are various approximation schemes available which provide a way
to cover this case as well. One crude approximation we shall adopt is to
consider the typical random graph chosen from a particular ensemble, and to
then further assume that this is well-approximated by just the average degree
of connectivity between an agent and its neighbors.
For the ensembles introduced earlier, i.e., for a $d$-dimensional hypercubic lattice
with some percolation, the average number of neighbors is:%
\begin{equation}
\text{Hypercubic Lattice: }n_{\text{avg}}=2d\times p_{\text{join}}%
\end{equation}
while for the Erd\"{o}s-Renyi ensemble, the average number of neighbors is:%
\begin{equation}
\text{Erd\"{o}s-Renyi: }n_{\text{avg}}=(M-1)\times p_{\text{join}},
\end{equation}
where $M$ is the total number of agents, i.e., nodes in the graph.

For hypercubic lattices, we can also introduce the notion of an effective dimension:
\begin{equation}
d_{\mathrm{eff}} = n_{\mathrm{avg}} / 2,
\end{equation}
a notion we shall also use (by abuse of terminology) for the Erd\"{o}s-Renyi ensemble as well.
With this in mind, we can reuse our previous analysis with a fixed connectivity, where
we replace all occurrences of $d$ by $d_{\text{eff}}$. In this case, there is
no need to confine our discussion to $d$ being an integer. When we turn to our
numerical experiments, we will indeed see that this approximation provides a
reasonable leading order characterization of the dynamics of branes.

\section{The Suburban Algorithm \label{sec:SUBURBAN}}

Having motivated the study of MCMC\ with strings and branes, we now turn to
some specific implementations of the suburban algorithm. For ease of
exposition, we shall present the case of sampling a single continuous variable
$x$. The generalization to a $D$-dimensional target (such as $\mathbb{R}^{D}$)
is straightforward, though there are various ways to do this, i.e.,
we can either adopt MH within a Gibbs sampler, or a sampler with joint variables (i.e., we perform
an update on all $D$ dimensions simultaneously).\footnote{To be more precise, the MH within Gibbs update for a target distribution
$p(x^{(1)},...,x^{(D)})$ with support on a $D$-dimensional space amounts to viewing this as a conditional probability
$p(x^{(1)},...,x^{(D)}) = p(x^{(i)} | x^{(1)},...,x^{(\widehat{i})},...,x^{(D)}) p(x^{(1)},...,x^{(\widehat{i})},...,x^{(D)})$, where the notation $\widehat{i}$ indicates that we omit this index. The MH within Gibbs update is then given by sampling from just the univariate distribution:
\begin{algorithm}[H]
\begin{algorithmic}
    \STATE Introduce $\Gamma = \{1,...,D\}$
    \STATE \textbf{for} $i = 1 \textrm{ to } D$ \textbf{do}
    \STATE \ \ \ \ $j \gets$ draw from $\Gamma$
    \STATE \ \ \ \ $x^{(j)} \gets$ sample from $p(x^{(j)} | x^{(1)},...,x^{(\widehat{j})},...,x^{(D)})$ using a 1D MH update.
    \RETURN $(x^{(1)},...,x^{(D)})$
\end{algorithmic}
\caption{MH within Gibbs}
\label{alg:MHGibbs}
\end{algorithm}
Both types of samplers have their relative merits, and we
will study examples of both.}

Let us now turn to the structure of the suburban sampler.
Recall that we are interested in a class of Metropolis-Hastings algorithms in
which instead of directly sampling from $\pi(x)$, we introduce
multiple copies of the target and sample from the joint distribution:%
\begin{equation}
\pi(x_{1},...,x_{M})=\pi(x_{1})...\pi(x_{M}).
\end{equation}
We shall also refer to the proposal kernel as:%
\begin{equation}
q(x_{1},...,x_{M}|y_{1},...,y_{M}, A)=\underset{\sigma=1}{\overset{M}{%
{\displaystyle\prod}
}}q_{\sigma}(x_{\sigma}|\text{Nb}(y_{\sigma})),
\end{equation}
where $A$ is the adjacency matrix of the grid.
To avoid overloading the notation, we shall write $\mathcal{X}^{(t)}\equiv
\left\{  x_{1}^{(t)},...,x_{M}^{(t)}\right\}  $ for the current state of the grid.
In what follows, we write the MH\ acceptance probability as:%
\begin{equation}
a\left(\mathcal{X}^{\text{new}}| \mathcal{X}^{\text{old}}, A \right)
=\min\left(  1,\frac{q(\mathcal{X}^{\text{old}}|\mathcal{X}^{\text{new}}, A)}{q(\mathcal{X}^{\text{new}}|\mathcal{X}^{\text{old}}, A)}
\frac{\pi\left(  \mathcal{X}^{\text{new}}\right)  }{\pi\left(\mathcal{X}^{\text{old}}\right)}\right).
\end{equation}
We now introduce algorithm \ref{alg:suburban}, the suburban algorithm.

\begin{algorithm}[t!]
\begin{algorithmic}
    \STATE Randomly Initialize $\mathcal{X}^{(0)}$ and $A^{(0)}$
    \STATE \textbf{for} $t = 0 \textrm{ to } N-1$ \textbf{do}
    \STATE \ \ \ \ $\mathcal{X}^{(\ast)} \gets$ sample from $q(\mathcal{X} | \mathcal{X}^{(t)}, A^{(t)})$
    \STATE \ \ \ \ accept with probability $a(\mathcal{X}^{\ast} | \mathcal{X}^{(t)}, A^{(t)})$
    \STATE \ \ \ \ \ \ \ \ \textbf{if} accept = true \textbf{then}
    \STATE \ \ \ \ \ \ \ \ \ \ \ \ $\mathcal{X}^{(t+1)} \gets \mathcal{X}^{(\ast)}$
    \STATE \ \ \ \ \ \ \ \ \textbf{else}
    \STATE \ \ \ \ \ \ \ \ \ \ \ \ $\mathcal{X}^{(t+1)} \gets \mathcal{X}^{(t)}$
    \STATE \ \ \ \ $A^{(t + 1)} \gets$ draw from $\mathcal{A}$
    \RETURN $\mathcal{X}^{(1)},...,\mathcal{X}^{(N)}$
\end{algorithmic}
\caption{Suburban Sampler}
\label{alg:suburban}
\end{algorithm}
An important feature of the suburban algorithm is that some of these steps can be parallelized whilst retaining detailed balance.
For example we can pick a coloring of a graph and then perform an update for all nodes of a particular color whilst holding fixed the
rest.

There are of course many variations on the above algorithm. For example, in
practice for each time step we shall perform a Gibbs update over our $M$ agents. For Gibbs sampling
over the target, we then have a Gibbs update schedule with $D \times M$ steps, and for the joint sampler,
it is over just $M$ steps. We can also choose to not draw a new
random graph $A^{(t)}$ at each step, but rather only every $T_{\text{draw}}$
steps. Other possibilities include stochastic time evolution for $A^{(t)}$.
To keep the analysis tractable, however, we will indeed stick to the simplest possibility,
performing an update on the graph topology at each sampling time step.

Now, having collected a sequence of values $\mathcal{X}^{(1)},...,\mathcal{X}%
^{(N)}$, we can interpret this as $N\times M$ samples of the original
distribution $\pi(x)$. As standard for MCMC\ methods, we can then calculate
quantities of interest such as the mean and covariance for the distribution
$\pi(x)$ by performing an appropriate sum over the observables:%
\begin{align}
\left\langle x\right\rangle _{\pi} &  \simeq\frac{1}{NM}\underset{\sigma,t}{%
{\displaystyle\sum}
}x_{\sigma}^{(t)}\\
\left\langle \left(  x-\left\langle x\right\rangle _{\pi}\right)
^{2}\right\rangle _{\pi} &  \simeq\frac{1}{NM-1}\underset{\sigma,t}{%
{\displaystyle\sum}
}\left(  x_{\sigma}^{(t)}-\left\langle x\right\rangle _{\pi}\right)  ^{2}%
\end{align}
as well as higher order moments.

Let us discuss the reason we expect our sampler to converge to the correct
posterior distribution. First of all, we note that although we are modifying
the proposal kernel at each time step (i.e., by introducing a different
adjacency matrix $A\in\mathcal{A}$), this modification is independent of the
current state of the system. So, it cannot impact the eventual posterior
distribution we obtain. Second, we observe that since we are just performing a
specific kind of MH\ sampling routine for the distribution $\pi(x_{1}%
,...,x_{M})$, we expect to converge to the correct posterior distribution.
But, since the variables $x_{1},...,x_{M}$ are all independent, this
is tantamount to having also sampled multiple times from $\pi(x)$.
The caveat is that we need the sampler to actually wander around during its random walk; $d \leq 1$ is typically necessary to prevent ``groupthink.''

\subsection{Implementation}

We now turn to the implementation of the suburban algorithm we shall
consider in subsequent sections. To accommodate a flexible framework for
prototyping, we have implemented the suburban algorithm in the probabilistic
programming language \texttt{Dimple} \cite{Dimple}. This consists of a set of
\texttt{Java} libraries with a \texttt{Matlab} wrapper. We have found this
interface to be quite helpful in reaching the form of the algorithm presented
in this work, as well as in performing different types of numerical experiments.

In the actual implementation, we have found it helpful to exclude some initial
fraction of the samples, i.e., the process known as \textquotedblleft
burn-in.\textquotedblright We do this more for practical considerations
connected with the diagnostics we perform than for any theoretical reason,
since a sufficiently well-behaved MCMC sampler run for long enough will eventually
converge anyway to the correct posterior distribution. In practice, we take a
fairly large burn-in cut, discarding the first $10\%$ of samples from a run, i.e., we only
keep $90\%$ of the samples. We always perform Gibbs sampling over the $M$ agents. If we also perform
Gibbs sampling over a $D$-dimensional target, we thus get a Gibbs schedule with $D \times M$ updates for each time
step. For a joint sampler, the Gibbs schedule consists of just $M$ updates.

The specific choice of proposal kernel we take is motivated by the physical considerations
outlined in section \ref{sec:MCEXTEND}:%
\begin{equation}
q_{\sigma}(x_{\sigma}|\text{Nb}(x_{\sigma}))\propto\exp\left(  -\alpha_{\sigma}\left(
D_{t}x_{\sigma}\right)  ^{2}-\underset{n(\sigma)}{\sum}\beta\left(  D_{t}x_{\sigma
} - D_{n(\sigma)}x_{\sigma}\right)  ^{2}\right)  \text{ \ \ with \ \ }%
\alpha_{\sigma}=2\beta-n_{\sigma}^{\mathrm{tot}}\beta,\label{propkernel}%
\end{equation}
that is, we take an adaptive value for the parameter $\alpha$ specified by the
number of nearest neighbors joined to $x_{\sigma}$. As already mentioned in section \ref{sec:MCEXTEND},
the main point is to ensure that the overall strength of the kinetic term, i.e., the quadratic terms
involving the temporal derivatives, does not dominate over the spatial derivatives.

In addition, we also implement the different choices of graph ensembles
outline in subection \ref{ssec:wgravity}. We also include the option to not
permute or \textquotedblleft shuffle\textquotedblright\ the indexing of the
agents. As a general rule of thumb, we find that switching off shuffling
always leads to worse performance.

\subsection{Hyperparameters}

Let us now formalize the total list of hyperparameters for the suburban
algorithm. The total number of timesteps is $N$, and the total number of
agents is $M$. In addition to the total number of samples collected in a run, we have a
choice of ensemble of random graphs, i.e., how we connect the agents together.
There is a coarse parameter given by the overall topology of graphs on which we perform percolation.
Additionally, we have introduced a class of ensembles where we permute the
locations of agents on the grid. For a collective of parallel MH\ samplers, this has no effect (since
there is no correlation between agents anyway), but for more general
collectives, this can clearly have an impact. Indeed, we find that if we
consider related ensembles in which shuffling is turned off, the
performance suffers. We shall therefore confine our
experiments to cases where shuffling is switched on. Finally, there is also a
continuous parameter $p_{\text{join}}$ which dictates the probability of a
given link in a graph being active. This in turn translates to the effective
worldvolume dimension experienced by an agent in the collective. Of course, the
specific choice of ensemble of random graphs will also affect how much
variance there is in the average degree of connectivity, though surprisingly,
this seems to be a subleading effect in the tests we perform.

There are also many hyperparameters lurking in the proposal kernel. For the
most part, we will focus on the case of equation (\ref{propkernel}), where
there is just one tunable parameter $\beta$. For a sampler in $D$ dimensions,
this naturally extends to a symmetric positive definite matrix $\beta_{IJ}$,
in the obvious notation. The overall parameter $\beta$ sets
the \textquotedblleft stiffness\textquotedblright\ or tension of the brane. For
$\beta$ large, the coupling to nearest neighbors is strongest, and the
relative size of jumps in the target space is smaller. For small $\beta$, the
brane is \textquotedblleft floppy,\textquotedblright\ and each agent in the
collective will execute larger movements. In this limit, the overall behavior
of the proposal kernel approaches the uniform distribution, and the effects
of grid topology are expected to become weaker.

\section{Overview of Numerical Experiments \label{sec:EXPERIMENT}}

Our emphasis up to this point has been on various theoretical aspects of the
suburban algorithm, in particular, how to understand MCMC with extended
objects. We now switch gears from theory to experiment, and ask how well such
algorithms do in practice. Our plan in this section will be
to give a list of the various metrics we shall use to gauge performance.
We then discuss the class of samplers we shall study, and then
give a brief overview of the target distributions we consider.
In subsequent sections we turn to examples.

For simplicity, we focus on the specific case where the
$q_{\sigma}$ of equation (\ref{Qextend}) are all normal distributions in which the
means and covariance matrix are dictated by the choice of nearest neighbors.
In most cases, we consider MH within Gibbs sampling, though we also
consider the case where joint variables are sampled, that is, pure MH.
For target distributions we focus on low-dimensional examples of target distributions such as various
mixture models of normal distributions, as well as the Rosenbrock \textquotedblleft
banana distribution,\textquotedblright\  which has most of its mass
concentrated on a lower dimensional subspace.

Rather than perform error analysis within a single long MCMC run, we opt to take multiple independent trials of
each MCMC run in which we vary the hyperparameters of the sampler such as the
overall topology and average degree of connectivity of the sampler. Though this leads to
more inefficient statistical estimators for our MCMC runs, it has the virtue of allowing us to easily
compare the performance of different algorithms, i.e., as we vary the continuous and discrete
hyperparameters of the suburban algorithm.

To gauge performance of the different runs, we focus on examples where we can
analytically compute various statistics such as the mean and covariance matrix
of the target distribution, comparing with the value obtained from our MCMC
samplers. We also calculate the expected number of samples on a tail to see
whether the sampler spends the correct amount of time searching for ``rare
events.'' We also collect the rejection rate and the integrated
auto-correlation time (i.e., mixing rate) for the MCMC sampler.

\subsection{Performance Metrics}

In general, gauging the performance of an MCMC\ algorithm can be difficult, so
we shall adopt a few different performance metrics. To keep the size
of the tests manageable (i.e., on the order of
a few weeks rather than months or years) we also limit the dimension of the
target space distributions we consider.

The performance metrics we adopt can roughly be split into tests of how well
the algorithm converges to the correct posterior distribution, i.e., external
comparisons, as well as internal comparisons such as the mixing rate and
rejection rate for the Markov chain.

Let us now discuss each of these performance metrics in more detail. As a
partial characterization of convergence to the correct posterior distribution,
we focus on target distributions where we can calculate the
various moments of the probability distribution analytically. In particular,
for a $D$-dimensional target, we obtain a sample value for the mean and
covariance matrix which we denote as $\mu_{\text{inf}}$ and
$\Sigma_{\text{inf}}$, respectively, i.e., the inferred values.
We then compute the distance to the true mean and covariance using the metrics:%
\begin{equation}
d_{\text{mean}}   \equiv\left\Vert \mu_{\text{inf}%
}- \mu_{\text{true}}\right\Vert \,\,\, \text{and} \,\,\,
d_{\text{cov}}  \equiv\left(  \text{Tr}\left(  \left(  \Sigma_{\text{inf}%
}-\Sigma_{\text{true}}\right)  \cdot\left(  \Sigma_{\text{inf}}-\Sigma
_{\text{true}}\right)  ^{T}\right)  \right)  ^{1/2},
\end{equation}
in the obvious notation. In addition to these simple tests, we also divide up
the distribution into various regions of high and low probability mass, and
verify that we obtain the appropriate number of events in each of these
regions.\ In practice, we always consider drawing a box of size $L^{D}$
centered at the origin such that there is a $68\%$ chance of falling inside
the box. We then also calculate the related $2\sigma$ and $3\sigma$ boxes
(respectively $95\%$ and $99.7\%$), and verify that a similar number of counts
falls in the appropriate bin. In practice, we actually concentrate on the
number of counts in the $0\sigma-1\sigma$ region, the $1\sigma-2\sigma$
region, the $2\sigma-3\sigma$ region, and events which fall outside the
$3\sigma$ region. For each such region, we compute the expected number of
events with the observed number, and obtain a corresponding fraction:%
\begin{equation}
f_{\text{region}}\equiv\frac{N_{\text{inf}}-N_{\text{true}}}{N_{\text{total}}%
},
\end{equation}
where $N_{\text{total}}$ denotes the total number of samples (after taking
into account burn-in).

In addition to these external metrics, i.e., metrics based on comparison with
the actual distribution, we also use diagnostics that are available from the
MCMC\ runs. These are important in most actual applications of MCMC since we do not usually
know the analytic form of the target distribution. Rather, we must depend on internal
diagnostics such as the rejection rate, and the integrated auto-correlation time, i.e., the mixing rate.
A typical rule of thumb is that for targets with no large free energy barriers, a rejection
rate of somewhere between $50\%-80\%$ is acceptable
(see e.g., \cite{Gelman:1997}). Indeed, if the rejection rate is too
low, then the sampler is wandering aimlessly, and if the rejection rate is too
high, it is an indication that too little of the target is being
explored. Let us also note, however, that in situations where there is a large
free energy barrier (i.e., multiple high mass regions separated by large low mass
regions), the rejection rate can turn out to be rather high. This is just a
symptom of the fact that most proposals will land in a low mass region.

Finally, we also collect the value of the integrated auto-correlation time for
the \textquotedblleft energy\textquotedblright\ of the distribution. This
observable provides a way of quantifying the correlation between samples drawn
at different times from the MCMC\ run. This observable in particular has been
argued to provide a preferred diagnostic for evaluating performance of an
MCMC\ run (for a recent discussion, see e.g., \cite{MCHammer}).
Along these lines, we introduce:
\begin{equation}
V=-\log \pi(x_{1},...,x_{M}),
\end{equation}
and collect the values $V^{(1)},...,V^{(N)}$. We evaluate the
covariance $c(k)$ for $-N<k<N$,%
\begin{equation}
c(k)\equiv\left\{
\begin{array}
[c]{c}%
\frac{1}{N}\underset{t=1}{\overset{N-k}{%
{\displaystyle\sum}
}}\left(  V^{(t)}-\overline{V}\right)  \left(  V^{(t+k)}-\overline{V}\right)  \text{
\ \ for \ \ }k\geq0\\
\frac{1}{N}\underset{t=1}{\overset{N+k}{%
{\displaystyle\sum}
}}\left(  V^{(t)}-\overline{V}\right)  \left(  V^{(t-k)}-\overline{V}\right)  \text{
\ \ for \ \ }k<0
\end{array}
\right\}  ,
\end{equation}
and then extract the cross correlation $\widehat{c}(k)$:%
\begin{equation}
\widehat{c}(k)=c(k)/c(0).
\end{equation}
From this, we extract an estimate for the integrated auto-correlation time:%
\begin{equation}
\tau_{\text{dec}}\equiv\underset{-N<k<N}{\sum}\left(  1-\left\vert \frac{k}%
{N}\right\vert \right)  \left\vert \widehat{c}(k)\right\vert
=1+2\underset{k=1}{\overset{N-1}{\sum}}\left(  1-\frac{k}{N}\right)
\left\vert \widehat{c}(k)\right\vert
\end{equation}
we also refer to this as the \textquotedblleft decay time,\textquotedblright
as it reflects how quickly the chain mixes. An important aspect of this
analysis is that we explicitly include all samples here, i.e., we do not
discard any samples from burn-in. When $\tau_{\text{dec}}$ is high, it means
our samples are highly correlated. To obtain a reliable estimate from an MCMC
run, it is then necessary to either perform thinning, i.e. only take some sparse fraction of
the original samples, or to run the sampler for even longer.

Now, for each of these observables, we could in principle extract standard
errors using just a single run of the MCMC\ algorithm. We could
also consider various sophisticated measures of convergence to the
correct posterior distribution (see e.g., \cite{RafteryLewis}).
Since we have analytic control over the
target distribution, we shall instead adopt
a somewhat cruder approach to such error bar estimates: We
simply repeat each experiment multiple times and collect both the means and
standard errors on our observables. We do this primarily in order to not bias
our analysis of errors which might otherwise depend on details of the
algorithm. Along these lines, we perform $T$ independent trials with random
initialization for each agent on $[-100, + 100 ]^{D}$. We present all plots with a $3$-sigma level standard
error around the mean value from these trials. In practice, we
typically find acceptable error bars for $T = 100$ and $T = 1000$ trials.

\subsection{A Sampling of Samplers}

Our primary interest is in the performance of the suburban sampler as we vary the different hyperparameters such as the grid
topology; the split / join rate; and the tension $\beta$ of the extended object. Now, in the special limit where we switch off
all links between agents, we get a collection of $M$ parallel MH samplers. We shall therefore
interchangeably use the notation $d_{eff} = 0$ and ``parallel MH'' samplers. Observe that this model also depends on the hyperparameter $\beta$ which controls the size of a proposed jump.

To gauge whether our performance is competitive with other simple
examples of samplers, we also compare the performance of the
suburban algorithm with slice sampling \cite{SliceSampler} within Gibbs. To make
a more direct comparison with parallel MH, we also run $M$ parallel slice samplers.

Even so, our goal is not to directly compare performance.
There are a few general issues with doing so. In slice sampling,
the algorithm typically queries the target distribution several times before recording a new sample value. In our comparison tests,
we always keep fixed the total number of samples. So, whereas MH always evaluates the target distribution
precisely twice on each loop (the new proposed value and the old value in the accept/reject step), the slice sampler will always make
more evaluations of the target distribution. In practice, we find that the number of evaluations can be a factor
of $5 \sim 10$ more when compared with MH. Since we have also not attempted to optimize the performance of a given algorithm,
a direct comparison with either CPU time or ``clock on the wall'' time would also seem premature.

Even so, the crude comparisons we do perform point to the fact that for
suburban samplers with stringlike connectivity $d_{eff} \sim 1$,
the overall performance is comparable to parallel slice samplers.  For additional details on
our implementation, and some example comparison runs, see Appendix \ref{app:SLICE}.

\subsection{Example Targets}

Let us now turn to the class of target space distributions we will use to test
our suburban samplers. In general, there are of course many possible choices
to make. Our examples are motivated primarily by the condition that we can
easily track the effect of changes in the various hyperparameters.

The largest class of examples we consider are Gaussian mixture models in $D$
dimensions. These consist of $k\geq1$ components with a set of means
$\mu^{(1)},...,\mu^{(k)}$ and covariance matrices $\Sigma^{(1)},...,\Sigma
^{(k)}$ so that the full target distribution is given by a weighted sum of
normal distributions:%
\begin{equation}
\pi_{\text{GMM}}(x)=\underset{l=1}{\overset{k}{\sum}}\,p^{(l)}\mathcal{N}%
(x|\mu^{(l)},\Sigma^{(l)}).
\end{equation}
Given sufficiently many experiments, we can tune the hyperparameters of an
MCMC\ routine to optimize a given sampler. In our experiments, we will vary
the number of mixtures, as well as the relative spacing between means and the
alignment of the covariance matrices. Since our aim is to compare the
performance of different samplers, we shall typically focus on models where we
do not need to fine-tune the hyperparameters of the model to get reasonably
accurate results.

As another class of examples, we also consider a case used in the study of
optimization routines known as the two-dimensional Rosenbrock \textquotedblleft banana
function\textquotedblright:
\begin{equation}
\pi_{\text{banana}}(x,y|\mu,\alpha) \propto \exp(-(x-\mu)^{2}-\alpha(y-x^{2})^{2}).
\end{equation}
In the literature, it is common to take $\mu=1$ and $\alpha=100$, i.e., even
though we have a two-dimensional target space, the high probability mass
region is localized along the one-dimensional subspace $y = x^2$. In such situations, we
can expect Gibbs sampling routines to fare poorly (a fact we verify), but
MH\ samplers with joint variables still provide a way to accurately sample from such distributions.

As a final comment, to keep the timescale of all experiments short, we have
also chosen to keep the total number of target space dimensions small, i.e., we
focus on $D=2$ and $D=10$. We expect that at least for the Gibbs samplers, our
conclusions continue to persist in higher dimensions. In the case of a joint
sampler we can expect some decrease in performance at large dimension (a not uncommon issue in MCMC).

\section{Effective Connectivity and Symmetric Mixtures \label{sec:DIMENSION}}

Perhaps the single most important feature of the suburban algorithm is that it
correlates the inferences drawn by nearest neighbors on a grid. Quite strikingly, we find
that the effective dimension rather than the overall topology of the grid
plays the dominant role in the performance of the algorithm.

To illustrate this general point, we will primarily focus on a simple example
in which we can isolate the effects of the different hyperparameters. We consider a
class of target distribution examples which we refer to as
\textquotedblleft symmetric mixtures.\textquotedblright For a fixed choice of
$D$ the number of target space dimensions, we introduce a mixture model
consisting of $2D$ components, with weights, means and covariance matrices:%
\begin{equation}
p^{(\pm,l)}=\frac{1}{2D}\text{, \ \ }\mu_{i}^{( \pm ,l)}= \pm \mu\times\delta_{i}%
^{l}\text{, \ \ }\Sigma^{(\pm, l)}=\sigma^{2} \times \mathbb{I}_{D \times D},
\end{equation}
where $i,j=1,...,D$ are indices running over the target, $l=1,...,D$ indexes
half of the components, and $\mu$ and $\sigma$ are real numbers. Here,
$\delta_{i}^{l}$ is a Kronecker delta function and $\mathbb{I}_{D \times D}$ is the $D \times D$ identity matrix.
Since we find qualitatively
similar behavior for $D=2$ and $D=10$, we will primarily present the plots of
our tests for the $D=2$ runs. We also specialize to the case of target
parameters $\mu=1.5$ and $\sigma^{2}=0.25$. In this case, we have four equally
weighted components of our mixture model, and there is a free energy barrier
separating these centers.

As a first class of tests, we consider sampling with different topology grids for
$N=10,000$ timesteps, with each grid consisting of $M=81$ agents.
This particular number is chosen since it factors as $M=9^{2}=3^{4}$, i.e., we
can take a hypercubic lattice in both one, two and four dimensions. Additionally,
we compare the performance with that of the Erd\"os-Renyi ensemble of random
graphs. For each choice of hyperparameter, we perform $T = 100$ trials and use this to collect
central values as well as standard errors.

We have scanned the value of $\beta$ in steps of factors of $10$, and
find that the performance is better around $\beta=0.01$, so we focus
on this case. In all cases, we find that for the different samplers,
the values of the observables $d_{\text{mean}}$, $d_{\text{cov}}$ and $f_{\mathrm{region}}$ are
all small, thus indicating reasonable convergence to the correct posterior distribution.

There is, however, a marked difference in the mixing rate as we vary the split / join probability
for the ensemble. In figure \ref{twodcubetaus} we
display the values of the integrated auto-correlation time as a function of the effective
dimension dictated by the split / join rate for a given grid topology.
Quite striking is the universal behavior of
the samplers as a function of the effective dimension near $d_{\text{eff}}%
\sim1$, i.e., for connectivity similar to that of a string.
We also see that near $d_{\text{eff}}=0$, i.e., for parallel
MH\ samplers, we see much slower mixing rates. Additionally, once we go beyond
$d_{\text{eff}} \gtrsim 1$, we also see that the overall performance of the sampler
suffers.%

\begin{figure}[ptb]%
\centering
\includegraphics[
scale = 1.0, trim = 25mm 70mm 0mm 70mm
]%
{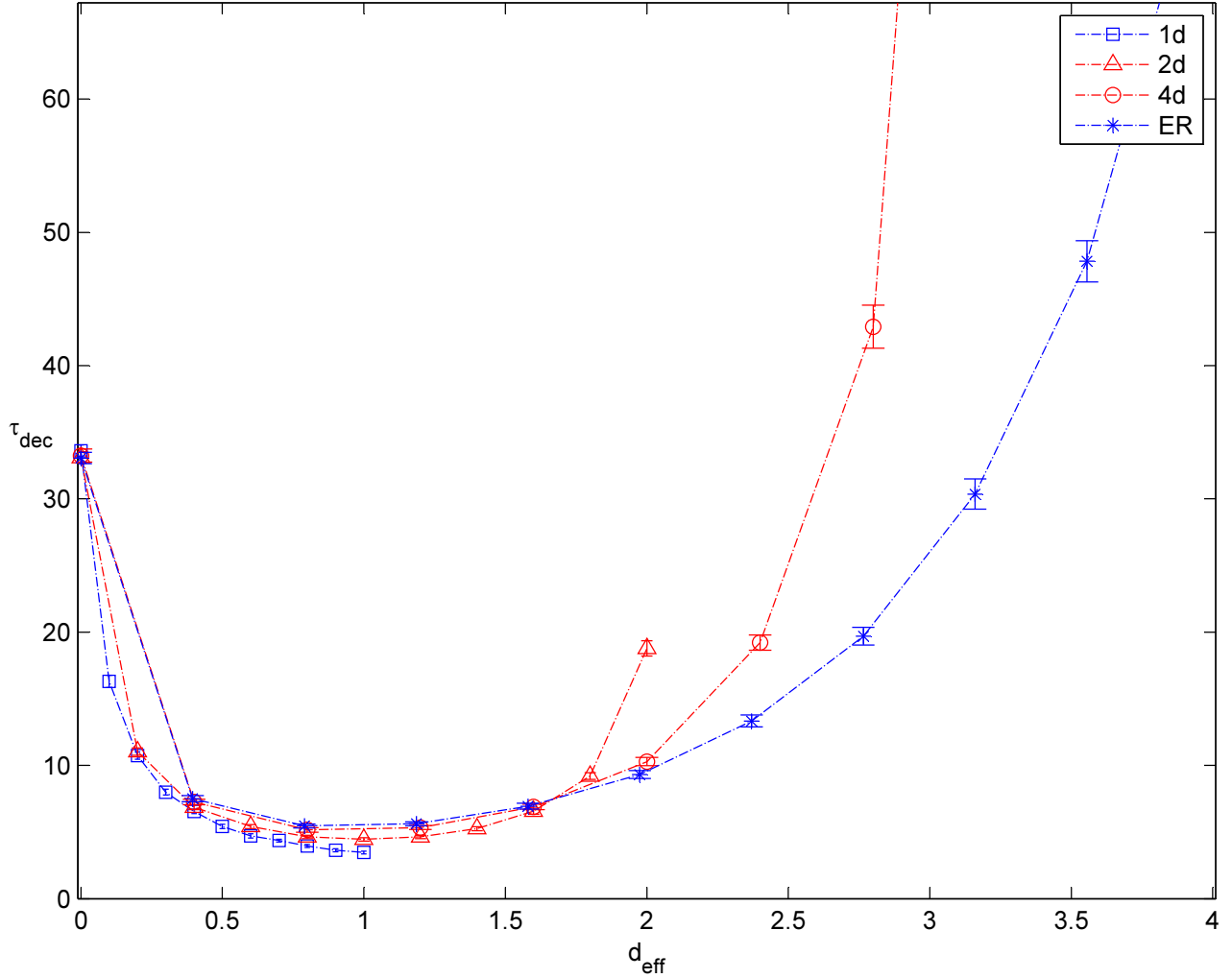}%
\caption{Plot of the integrated auto-correlation time $\tau_{\text{dec}}$ as a
function of the effective dimension of the grid. All data comes from sampling
the $D = 2$ symmetric mixture model. All runs are with $N=10,000$ timesteps and
$M=81$ agents with brane tension $\beta=0.01$, and with each hyperparameter subject to $T = 100$
independent trials. In the plot, the case of a grid
with topology a 1d, 2d and 4d hypercubic lattice, as well as the Erd\"os-Renyi ensemble
(ER) of random graphs are all plotted. Quite strikingly, there is a universal
behavior for $d_{\text{eff}}\lesssim 1$.}%
\label{twodcubetaus}%
\end{figure}

Due to this universal behavior which is independent of grid topology, we shall often
concentrate on \textquotedblleft representative behavior\textquotedblright\ as
obtained from a grid with topology that of a 2d membrane. In this case, when
an agent attaches to an average of two out of four nearest neighbors, we get
$d_{\text{eff}}=1$.

\begin{figure}[ptb]%
\centering
\includegraphics[
scale = 1.0, trim = 25mm 70mm 0mm 70mm
]%
{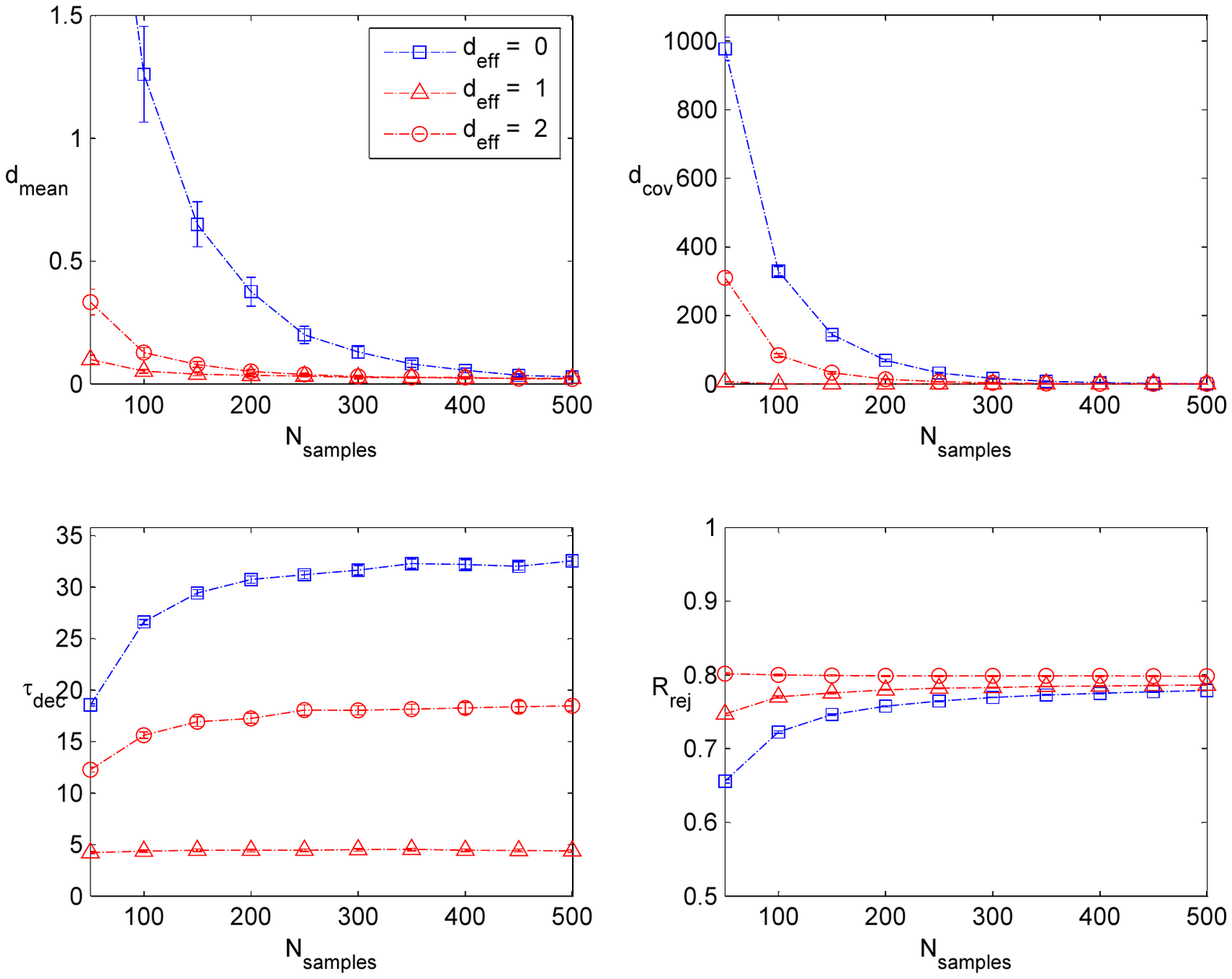}%
\caption{Plots of the distance to the true mean, true covariance, integrated
auto-correlation time and rejection rate as a function of the number of
timesteps for sampling from a 2D symmetric mixture model. Here, we vary the
total connectivity of the grid in a 2d membrane model with brane tension
$\beta=0.01$. For each choice of hyperparameter, we run for $T = 100$ independent trials.
Overall, $d_{\text{eff}}=0$ (i.e., parallel MH\ samplers) performs the worst, and the stringlike
$d_{\text{eff}}=1$ fares the best.}%
\label{timetwodcubemets}%
\end{figure}

\begin{figure}[ptb]%
\centering
\includegraphics[
scale = 1.0, trim = 25mm 70mm 0mm 70mm
]%
{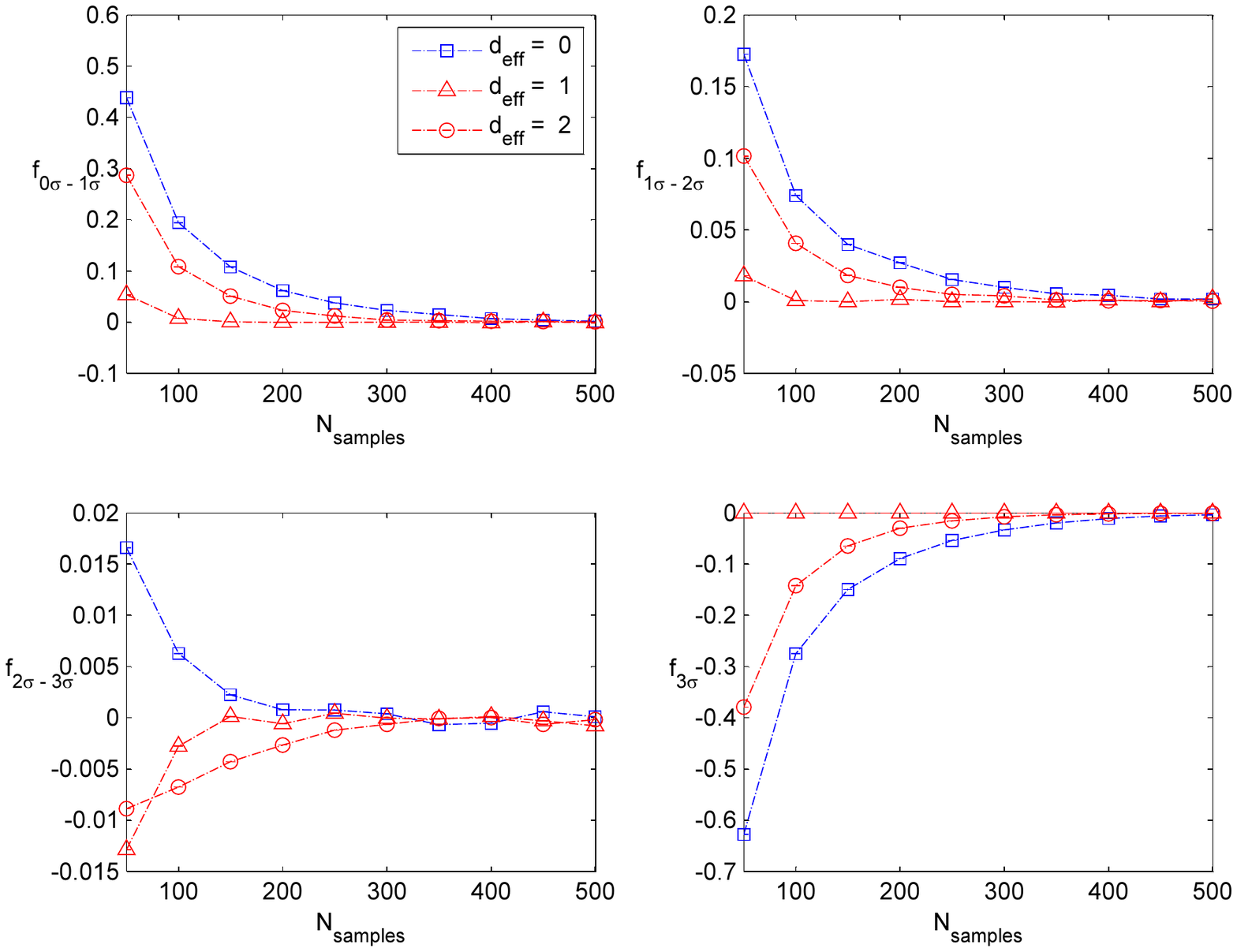}%
\caption{Plots of the fraction of total counts in the $1 \sigma$ interior region
(i.e., $68\%)$, in between the $1\sigma - 2 \sigma$ regions, $2\sigma - 3\sigma$ regions,
and outside the $3 \sigma$ region as a function of the total number of timesteps
taken by the sampler.}%
\label{timetwodcubetails}%
\end{figure}

To further probe the effects of the nearest neighbor interactions on the speed
of convergence, we have also looked at the performance as a function of the
number of samples. In figure \ref{timetwodcubemets} we show the speed of
convergence towards the true mean and covariance, as well as the behavior of
the integrated auto-correlation time and the rejection rate. Figure
\ref{timetwodcubetails} shows a similar collection of plots for the accuracy
of the total number of counts for various thresholds. Overall, we find that for
$d_{\text{eff}}\sim1$, this leads to comparable convergence rates. By
inspection, we also see that a collection of parallel MH samplers is far slower
in reaching an accurate inference.

The overall behavior provides a few general lessons consistent
with the theoretical arguments discussed earlier. First, we see that
the speed of convergence is dependent on the effective dimension
experienced by the agents. In particular, the mixing rates are best for
$d_{\text{eff}}\sim1$, i.e., for the case closest to a one-dimensional string.
Perhaps more surprising is the fact that even varying the topology of the grid
makes little difference at values $d_{\text{eff}}\lesssim1$.

\section{Random Landscapes \label{sec:LANDSCAPE}}

In this section we consider a more general class of mixture models in which there is a
landscape of local maxima and minima. The
presence of the additional hills and valleys means that a compromise will need
to be struck between \textquotedblleft wandering freely\textquotedblright\ as
will happen in the case of small $d_{\text{eff}}\sim0$, and moving more slowly
around individual components of the mixture model, i.e., for $d_{\text{eff}}\gtrsim1$.

To study this class of possibilities, we construct a few randomly selected
examples of two-dimensional landscapes. We have used a variant on the same random mixture model considered
in reference \cite{Herding}.
We focus on the case of $20$ mixtures with relative weights randomly drawn from the uniform
distribution on $[0,1]$. For each component of the mixture, we draw a random
two-dimensional vector with components in $[-0.4,+0.4]^{2}$, and randomly
drawn covariance matrix given by:%
\begin{equation}
\Sigma=O^{T}\cdot\left[
\begin{array}
[c]{cc}%
19r_{1}+1 & \\
& 19r_{2}+1
\end{array}
\right]  \cdot O
\end{equation}
where $O$ is a $2\times2$ random orthogonal matrix, and the $r_{i}$ are random
numbers drawn from the uniform distribution on $[0,1]$.

For different choices of random seeds, we then get a random
mixture model. By design, we have chosen our domain for the random variables
so that the brane tension $\beta = 0.01$ should give a roughly comparable
class of length scales for the target distribution. Since the overall topology
of the grid does not appear to affect the qualitative behavior of the sampler,
we have also focussed on the case of a two-dimensional grid with some amount
of percolation to control the effective dimension, which for a two-dimensional
grid can range over the values $0\leq d_{\text{eff}}\leq2$. For each choice of hyperparameter,
we perform $T = 100$ independent trials.

\begin{figure}[ptb]%
\centering
\includegraphics[
scale = 1.0, trim = 25mm 70mm 0mm 70mm
]%
{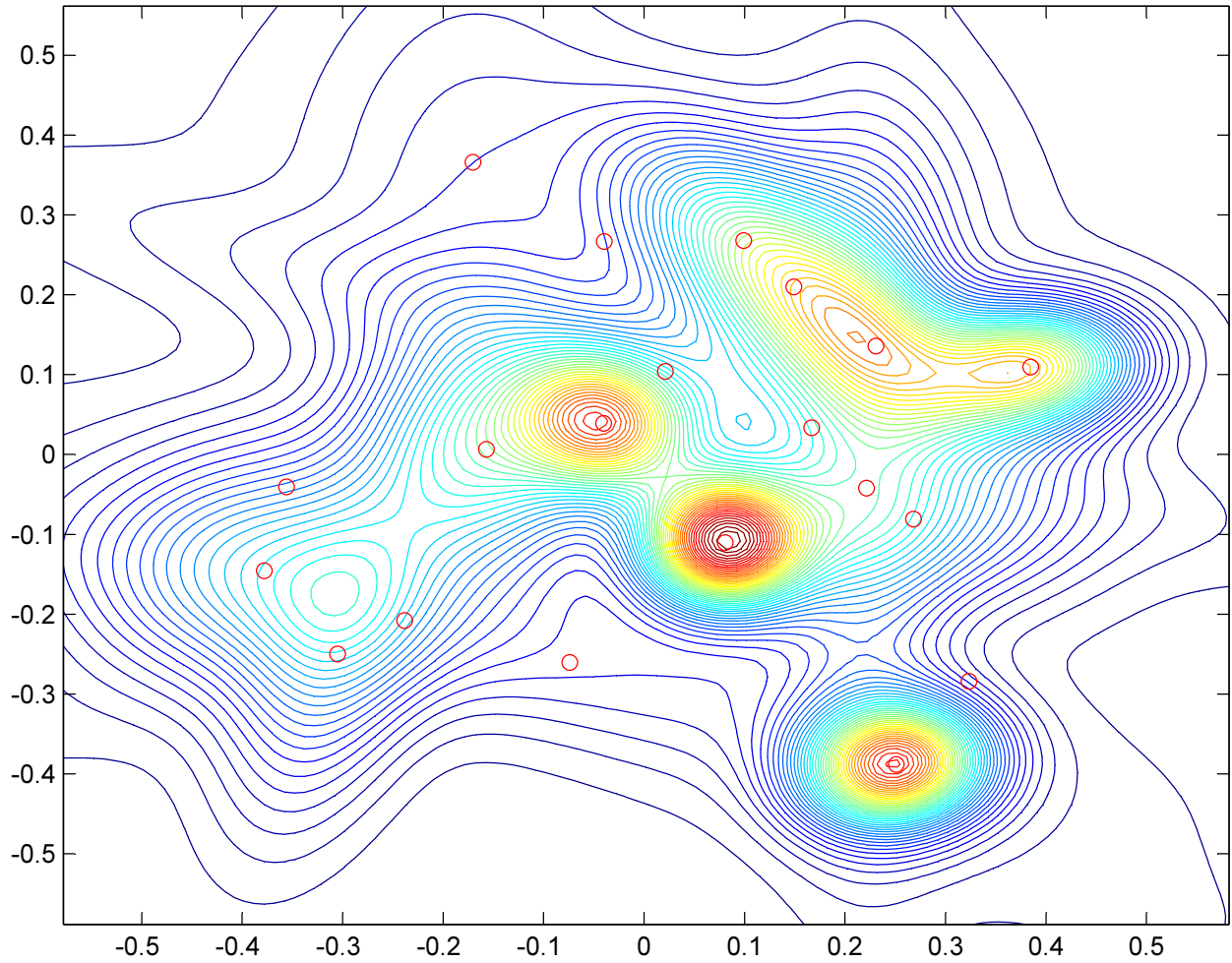}%
\caption{Contour plot of the random mixture model with twenty components
and random seed of $40$. Red circles denote centers of individual components.}%
\label{rand40plot}%
\end{figure}

\begin{figure}[ptb]%
\centering
\includegraphics[
scale = 1.0, trim = 25mm 70mm 0mm 70mm
]%
{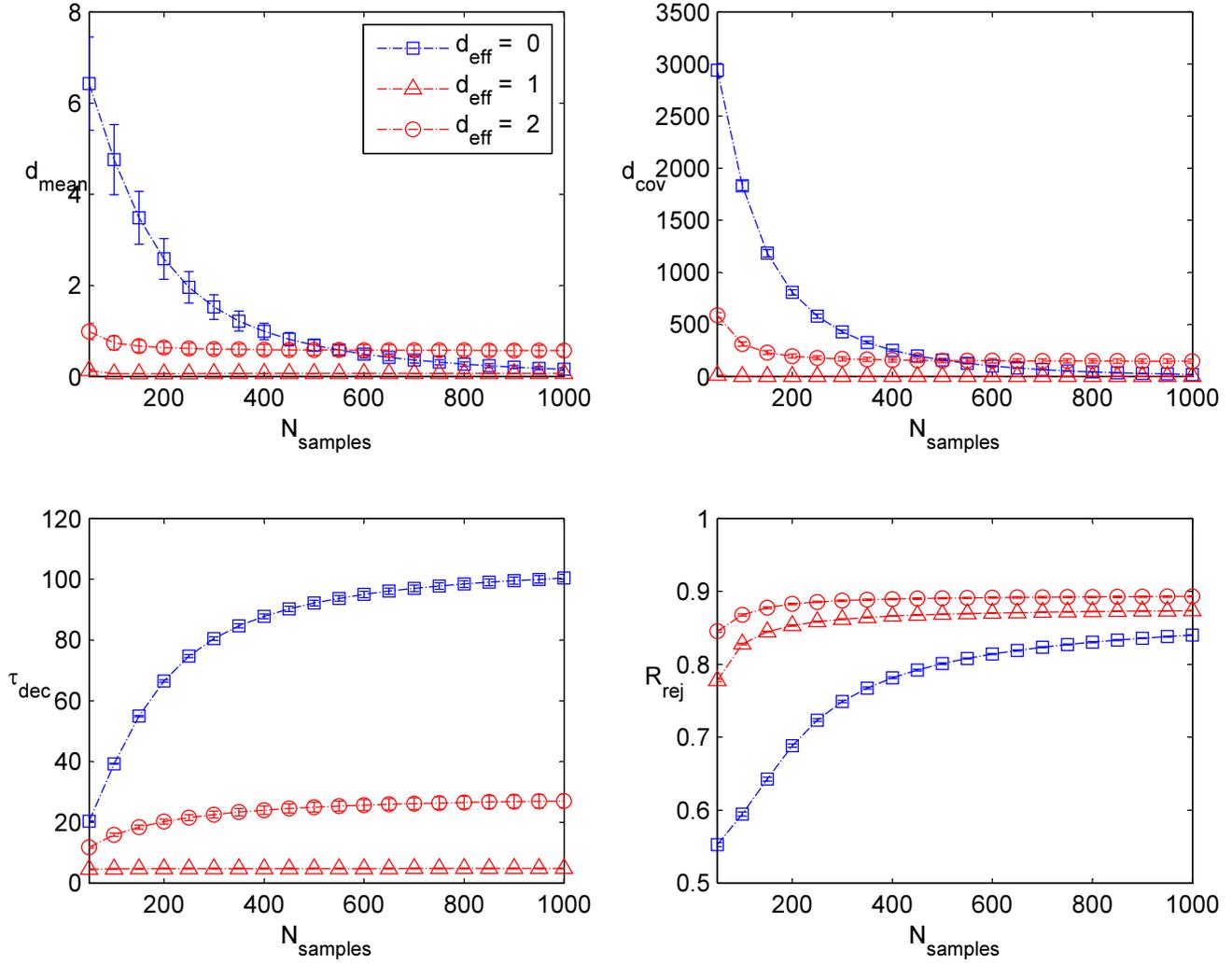}%
\caption{Plots of the convergence to the true mean and covariance matrix, as
well as integrated auto-correlation time and rejection rate for suburban
samplers of the random mixture model with random seed $40$. By
inspection, we observe that near $d_{\text{eff}}=1$, the samplers reach an
accurate inference more quickly than either the case of parallel MH\ samplers
($d_{\text{eff}}=0$), or a grid which is highly connected ($d_{\text{eff}}%
=2$).}%
\label{rand40b0p01mets}%
\end{figure}

\begin{figure}[ptb]%
\centering
\includegraphics[
scale = 1.0, trim = 25mm 70mm 0mm 70mm
]%
{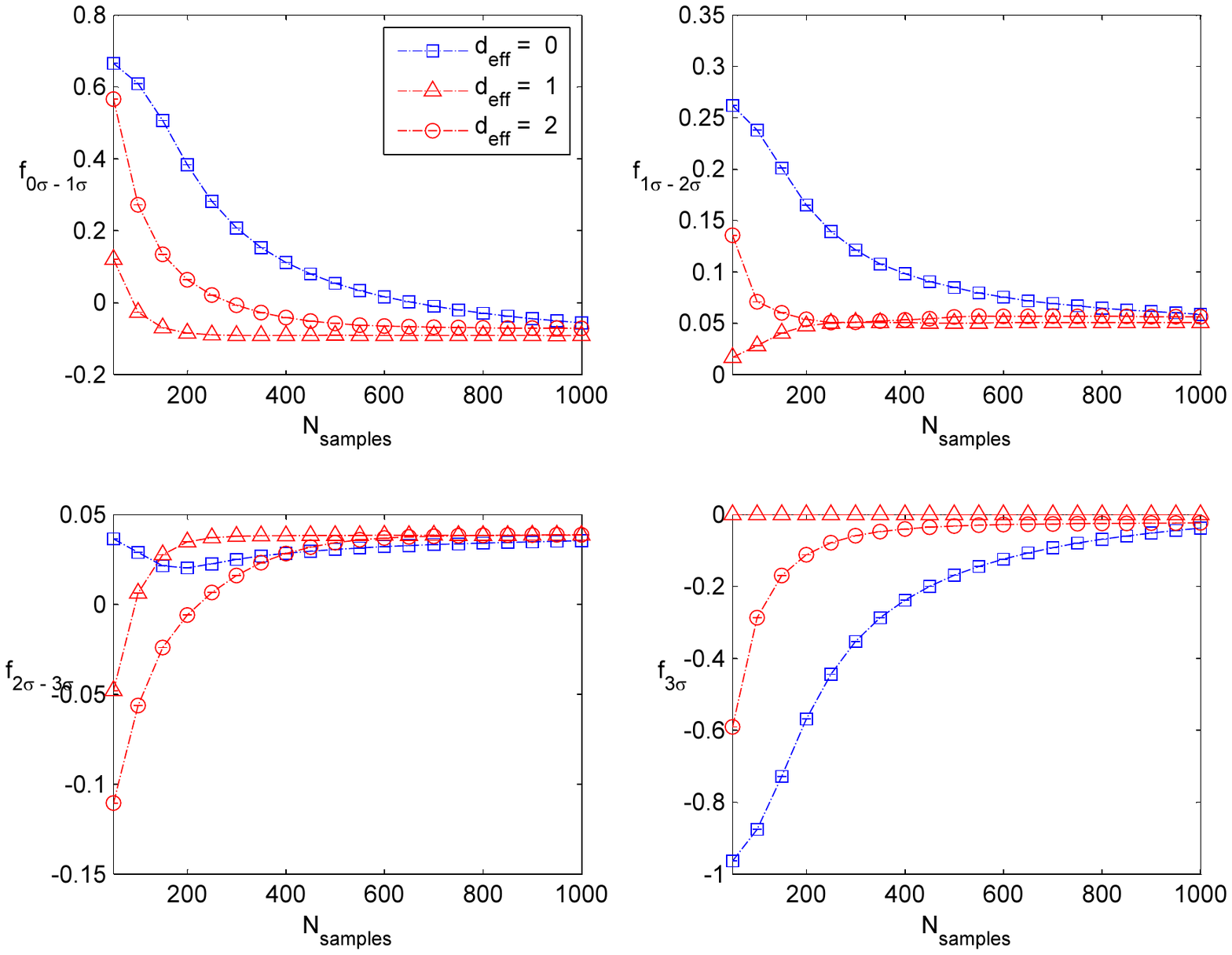}%
\caption{Plots of the convergence to the true tail statistics for suburban
samplers of the random mixture model with random seed $40$. By
inspection, we observe that near $d_{\text{eff}}=1$, the samplers reach an
accurate inference more quickly than either the case of parallel MH\ samplers
($d_{\text{eff}}=0$), or a grid which is highly connected ($d_{\text{eff}}%
=2$).}%
\label{rand40b0p01tails}%
\end{figure}

Though we have not exhausted the tests of possible randomly generated
landscapes, we have found that for the most part, there is an overall behavior
which is observed in the majority of examples. Compared with the random
mixture model of reference \cite{Herding}, we take the parameters \texttt{stdmu }%
$=0.4$, \texttt{stdsig} $=10.0$. We do this primarily to achieve convergence
for the samplers in a reasonable amount of time. The different mixture models
are obtained by setting the random seed in the code of \cite{Herding} to different
values. Figure \ref{rand40plot} shows a plot of the distribution with random seed set
to $40$. The performance of samplers in this case is similar to that observed for
the majority of runs. In figures \ref{rand40b0p01mets}\ and
\ref{rand40b0p01tails} we show plots of some of the performance metrics. In
particular, we see that the suburban sampler at $d_{\text{eff}} \sim 1$ exhibits a
faster and more accurate initial approach to the target distribution when
compared with parallel MH\ samplers (i.e., $d_{\text{eff}}=0$). We also observe
that when $d_{\text{eff}}=2$, the sampler settles in an
incorrect metastable configuration. We take this to mean that
\textquotedblleft groupthink\textquotedblright\ has developed within the
ensemble, a phenomena we generically expect when $d_{\text{eff}}$ becomes
larger than one. This appears to the generic behavior for such random mixture
models, though we have also observed some outlier behavior, for example when
the random seed is $5$.\footnote{In the case of the distribution generated by
the random seed $5$, we find that the mixing rate is again much faster for the
surburban sampler when compared with the parallel MH\ samplers. Nevertheless,
in this case, we find that the overall accuracy of the inference at later
times for parallel MH\ becomes comparable to that of the surburban samplers.} Finally, in some of the tail
statistics tests presented in figure \ref{rand40b0p01tails}, we observe that the sampler
is not actually converging to the correct value of the tail statistic. We take this to mean
that some agents in the ensemble have become stuck wandering in a metastable configuration. Note, however,
that even though the samplers with different effective dimension all appear to converge to this incorrect inference,
the suburban sampler reaches this conclusion (albeit incorrect) more quickly. It would
be interesting to understand this point further.

\section{Banana Distribution \label{sec:BANANA}}

A painful fact of life is that targets with most of their mass concentrated along a
low-dimensional subspace pose a challenge for some (untuned) samplers. It is therefore of interest to
study the performance of the suburban algorithm in such situations. In this section we forego Gibbs sampling and instead
focus on the case of a joint sampler, i.e., where we perform an update across all $D$
dimensions simultaneously.

We consider the special case of the
two-dimensional Rosenbrock probability density or \textquotedblleft banana
distribution\textquotedblright:%
\begin{equation}
\pi_{\text{banana}}(x,y)=\frac{1}{10\pi}\exp\left(  -(x-1)^{2}-100(y-x^{2}%
)^{2}\right)  .
\end{equation}
The key feature of this distribution is that the second term enforces the
approximate constraint $y \simeq x^{2}$, leading to an effectively
lower-dimensional distribution. The presence of this lower-dimensional ridge
is often used as a way to gauge the performance of optimization algorithms.

As already mentioned, we focus on a suburban sampler with joint variables.
We take different grid topologies for the statistical agents and then perform a
sweep over different values of the hyperparemeters $\beta$ and $d_{\text{eff}%
}$. For each choice of fixed hyperparameters, we then perform $100$ trials
where we initialize each agent with the uniform random distribution on $[-100,+100]^{2}$.%

\begin{figure}[ptb]%
\centering
\includegraphics[
scale = 1.0, trim = 25mm 70mm 0mm 70mm
]%
{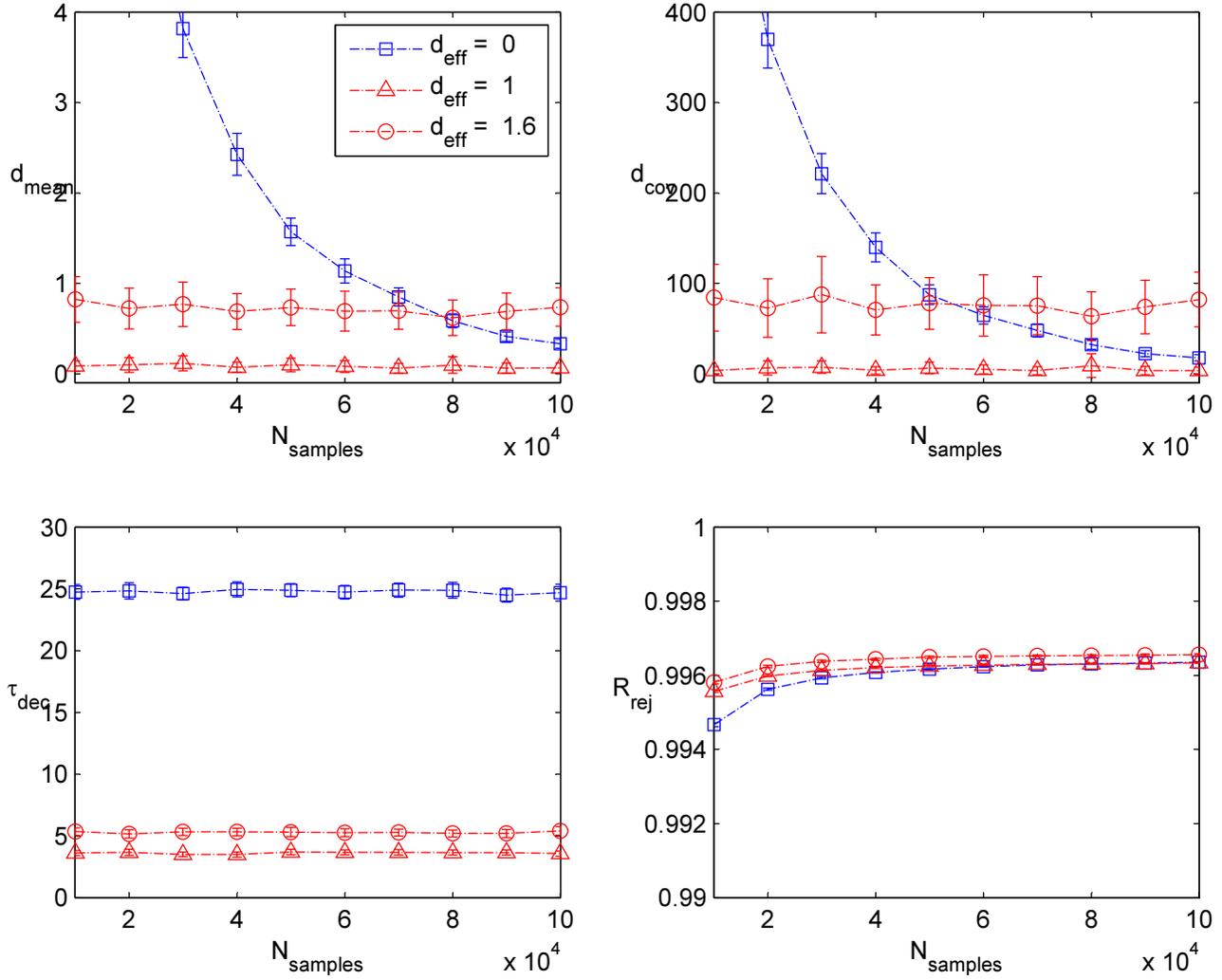}%
\caption{Plots of the convergence to the true mean and covariance matrix, as
well as integrated auto-correlation time and rejection rate for suburban
samplers of the banana distribution. By inspection, we observe that near
$d_{\text{eff}}=1$, the samplers reach an accurate inference more quickly than
either the case of parallel MH\ samplers ($d_{\text{eff}}=0$), or a grid which
is highly connected ($d_{\text{eff}}=1.6$).}%
\label{time2dbananamets}%
\end{figure}

Since we find qualitatively similar behavior for different grid topologies, we
shall focus on the representative case of a two-dimensional grid where we
allow splitting/joining and shuffling. We have also focussed on the special
case $\beta=0.01$. On general grounds, we expect this choice of hyperparameter
to fare better than other choices because in this special case it is tuned to
the dominant factor $100(y-x^{2})^{2}$ appearing in the exponential. We have
confirmed this point in the specific sweep over hyperparameters. We also let
the samplers run for a varying number of total samples ranging from
$N_{\text{samples}}=10^{4}$ to $N_{\text{samples}}=10^{5}$.

\begin{figure}[ptb]%
\centering
\includegraphics[
scale = 1.0, trim = 25mm 70mm 0mm 70mm
]%
{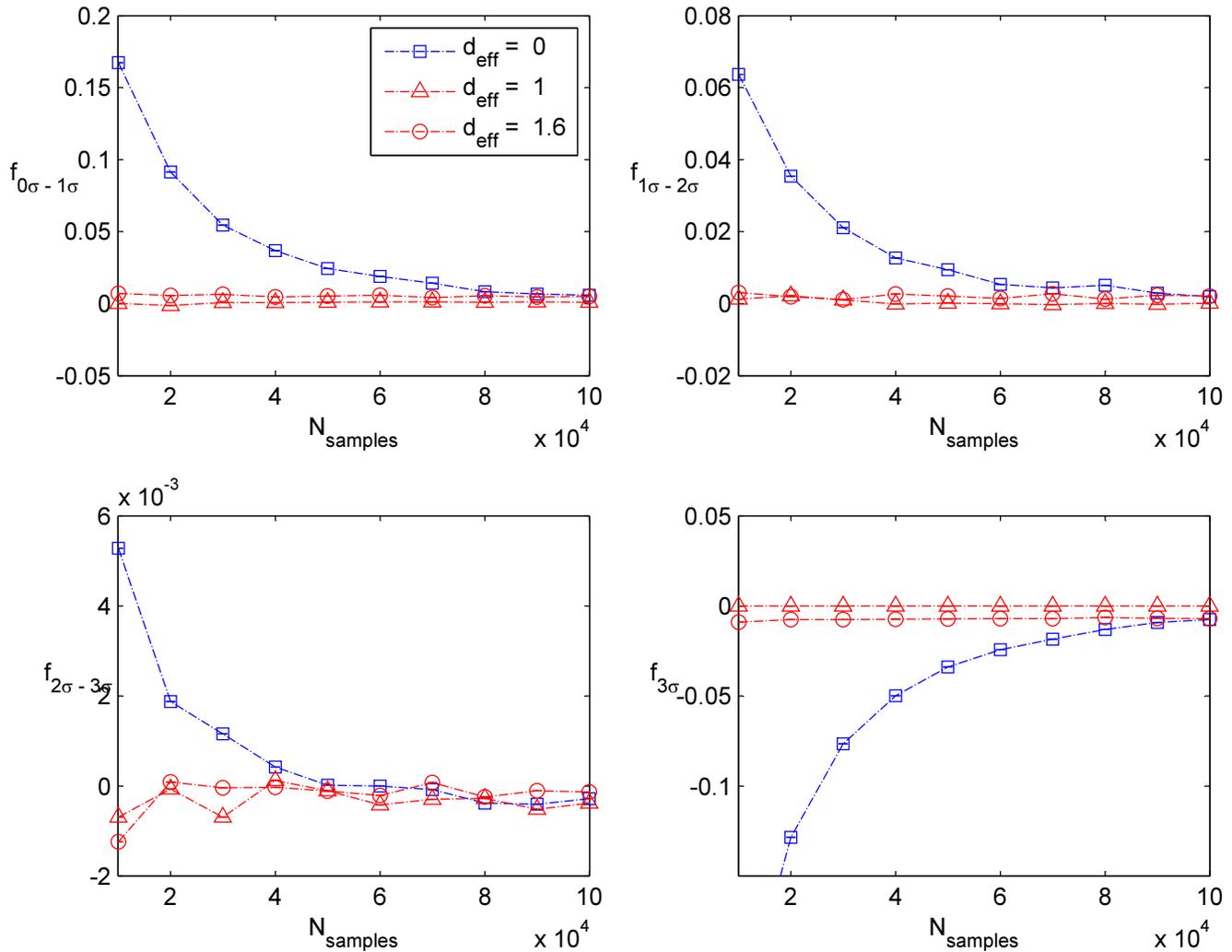}%
\caption{Plots of the convergence to the true correct tail statistics for
suburban samplers of the banana distribution. By inspection, we observe that
near $d_{\text{eff}}=1$, the samplers reach an accurate inference more quickly
than either the case of parallel MH\ samplers ($d_{\text{eff}}=0$), or a grid
which is highly connected ($d_{\text{eff}}=1.6$).}%
\label{time2dbananatails}%
\end{figure}

In figure \ref{time2dbananamets} we display the behavior of various
performance metrics as a function of the total number of samples collected. In
general, we find that the performance near $d_{\text{eff}}\sim1$ appears to
converge more quickly and accurately than that of a collection of parallel
MH\ samplers (i.e., $d_{\text{eff}}\sim0$). We also observe a similar
phenomenon noted in other target distributions: Once we pass to
$d_{\text{eff}}\gtrsim 1$, there can be a seemingly quick convergence to
\textquotedblleft an answer\textquotedblright\ though the actual accuracy of
this answer is difficult to correct due to the appearance of groupthink in the
ensemble. To illustrate this point, we show the behavior of the collective for
$d_{\text{eff}}=1.6$, where the case of parallel MH\ samplers eventually
overtakes the accuracy of the inferred mean and covariance matrix compared
with a highly connected grid (but not the case with $d_{\text{eff}}\sim1$).
Interestingly enough, we find that even in this situation the integrated
auto-correlation time for a connected grid provides a faster mixing rate
compared with a parallel MH\ sampler. This is in general accord with our
theoretical discussion presented earlier. Finally, the
rejection rate ---as expected--- is quite high, approaching order $0.996$. Part of
the reason for this high rejection rate is the presence of a narrow ridge
where all the probability mass is concentrated. Nevertheless, we can also see
that even though the rejection rate is high, the overall performance in the
other metrics, including the accuracy of counts on the tails of the target
(see figure \ref{time2dbananatails}) is also reasonably accurate.

\section{Free Energy Barriers \label{sec:FREEBALL}}

The extended nature of the suburban sampler suggests that for target distributions
with various disconnected deep \textquotedblleft pockets,\textquotedblright%
\ different pieces of the ensemble can wander over to different regions, and
thereby more efficiently reach a global characterization of the target. To
test this hypothesis, we have also considered a few examples of targets where
we vary the overall size of a possible free energy barrier.

\begin{figure}[t!]%
\centering
\includegraphics[
scale = 1.0, trim = 25mm 70mm 0mm 70mm
]%
{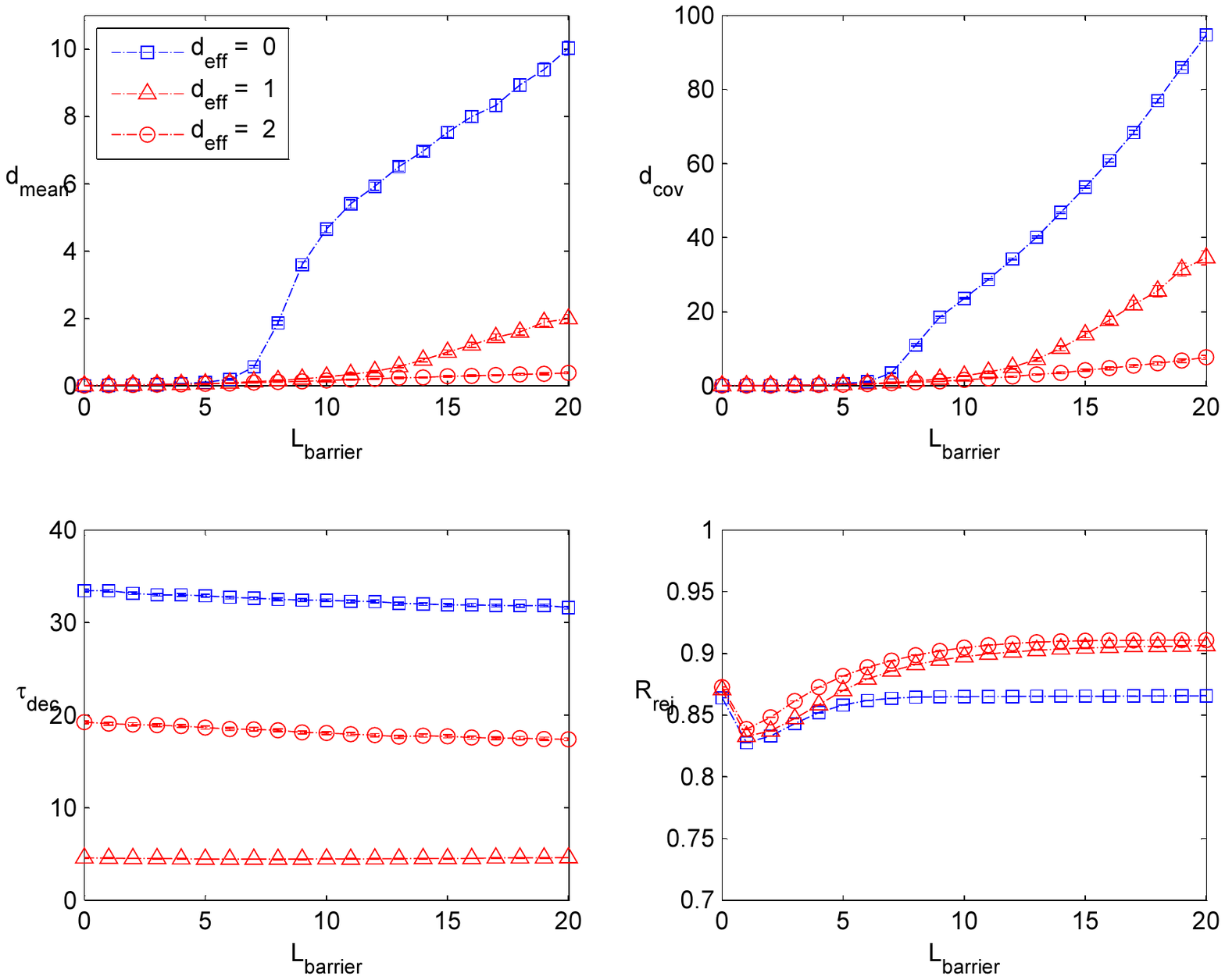}%
\caption{Plot of the convergence to the true mean, covariance matrix, as well
as the integrated auto-correlation time and rejection rate for suburban
samplers on a $D=2$ two component mixture model
where the centers are separated along the x-axis so that the total distance
between the two is $2L_{\text{barrier}}$.}%
\label{2dfreebarrierplotmet}%
\end{figure}

To avoid introducing too many extraneous parameters in our test, we focus on
the simple case of a target in $D$ dimensions with a two component normal
mixture model:%
\begin{equation}
\pi_{\text{GMM}}(x)=\frac{3}{4}\mathcal{N}(x|\mu^{(+)},\Sigma)+\frac{1}%
{4}\mathcal{N}(x|\mu^{(-)},\Sigma).
\end{equation}
We take unequal weights for the mixture model so that the mean is off-center from
the origin. This helps to ensure that the random initialization of values (which on average has mean at the zero)
does not accidentally align with the mean of the distribution.
Each component of the mixture model has the same covariance matrix,
but with the means separated along one axis:
\begin{equation}
\mu^{(+)}=(+L_{\text{barrier}},0,...,0)\text{, \ \ }\mu^{(-)}%
=(-L_{\text{barrier}},0,...,0)\text{, \ \ }\Sigma=\sigma^{2}\times
\mathbb{I}_{D\times D},
\end{equation}
where $L_{\text{barrier}}$ and $\sigma$ are the two numerical inputs to the target. In what
follows we always hold fixed $\sigma=0.25$, and vary the centers of the
Gaussians by changing $L_{\text{barrier}}$ from $0$ to $20$. To present a
uniform class of examples, we have taken $N_{\text{samples}}=1000$ samples
with $M=81$ agents on varying grid topologies. We primarily focus on
MH\ within Gibbs sampling where we take the tension to be $\beta=0.01$. For each choice of
hyperparameter, we perform $T = 1000$ independent trials.
Due to the fact that we are using a Gibbs sampler, we do not expect much
decrease in performance in comparing the $D=2$ and $D=10$ free energy barrier
tests, which we indeed verify. For the sake of brevity, we therefore
present the results from the $D = 2$ runs.

An interesting feature of this analysis is that as we increase the
distance between the centers of the Gaussians, the suburban samplers
tend to suffer less compared with their parallel sampler
counterparts. Additionally, the counts of the tail
statistics do not suffer as we increase the size of the free energy barrier.
See figure \ref{2dfreebarrierplotmet} for plots of some of the performance
metrics.

Note that for $d_{\text{eff}}=2$, the accuracy of the
inference is sometimes better than both parallel MH and
$d_{\text{eff}}=1$ samplers. In this situation, this is not
altogether surprising since a sampler suffering from groupthink will
nevertheless be able to find the \textquotedblleft hot
spots\textquotedblright\ in the distribution.

But the mixing rate for the $d_{\text{eff}}=2$
sampler is significantly slower than that of the $d_{\text{eff}}=1$ sampler. This means
that to get accurate estimators, we either need to perform thinning on the
$d_{\text{eff}} = 2$ samples (leading to worse performance), or run for longer.

\section{Conclusions \label{sec:CONC}}

In this paper we have introduced a physical picture for MCMC with extended objects. We have explained how for
an average connectivity with effective dimension $d_{\mathrm{eff}} \sim 1$, there are often benefits
to collective inference by an ensemble. Conversely, we have seen that ``groupthink'' can also set in at high connectivity.
We have also presented the results of various experimental tests of the suburban algorithm as a function of the overall topology
and splitting / joining rate. Quite strikingly, the key criterion which appears to affect performance is the average degree of connectivity, i.e., the number of nearest neighbors in the ensemble rather than the specific grid topology. We have also seen that when compared with parallel MH samplers, a suburban sampler with appropriate degree of connectivity has a faster mixing rate, as well as a more accurate convergence to the true
moments of the target. In the remainder of this section we discuss some potential future directions.

Clearly, there are a number of generalizations available which it would be interesting to explore further in
future work. A simple example of a generalization would be to study the effects of different proposal kernels.

Perhaps the single biggest change would be to implement a parallelized version of the updating schedule.
Indeed, one of the important features of the suburban algorithm is that the update steps for a given agent only depends on its nearest neighbors.
By a suitable blocking scheme, we can then consider multiple updates of the neighbors.

In general, we must exercise some caution in how many agents we simultaneously update, since we need to make sure that the priors of a given MH proposal obey detailed balance. To illustrate, let us focus on the special case where the topology of the grid is a $d$-dimensional hypercubic lattice. The lattice defines a bipartite graph, i.e., colored as black and white such that a vertex of one color always attaches to vertices of the other color. The parallelized algorithm which respects detailed balance is given by updating all of the agents on black vertices, and then performing an update on all the agents on white vertices. By construction, all of the neighbors of a given vertex are held fixed during a given update step.

In practice the loss of detailed balance may be acceptable, though it is then less clear whether we should expect convergence
to the correct posterior distribution. Though our implementation has focussed on the case of updating one agent at a time, we could also consider the opposite limit where all agents update simultaneously. Exploring this and related questions would be quite interesting.

Another direction which would be exciting to explore further is the generalization of the suburban sampler to targets with different data types. For example, another physically well-motivated class of targets involve discrete variables (e.g., Ising models).

Finally, aside from these more ``applied'' directions, it is tempting to turn the discussion
around to more fundamental issues and ask what lessons (even if preliminary) we can now draw for string theory and quantum gravity.
Indeed, though our implementation has taken certain liberties with the structure of the physical superstring,
we can see that there is nothing ``fundamental'' about the topology of a 1d or 2d grid. Rather, it is the effective number of neighbors which plays the crucial role in dictating the accuracy of an inference. Another point is that in the physical string, the effective potential
explored by a mobile string can be viewed as a background condensate of strings. It would be interesting to develop this point of view further.

\section*{Acknowledgements}

JJH thanks D. Krohn for collaboration at an early stage. We thank
J.A. Barandes, C. Barber, C. Freer, M. Freytsis, J.J. Heckman Sr., A. Murugan, P. Oreto, R. Yang, and J. Yedidia for helpful discussions.
JJH also thanks the theory groups at Columbia University, the ITS at the CUNY\ graduate center, and the CCPP at
NYU for hospitality during the completion of this work. The work of JJH is supported by NSF CAREER grant PHY-1452037.
JJH also acknowledges support from the Bahnson Fund at UNC Chapel Hill as well as the
R.~J. Reynolds Industries, Inc. Junior Faculty Development Award from the Office
of the Executive Vice Chancellor and Provost at UNC Chapel Hill.
JJH also thanks the virtual computing lab (VCL) at UNC Chapel Hill for assistance.

\appendix

\section{Comparison with Slice Sampler} \label{app:SLICE}

In this Appendix we present some additional details on our comparison tests with slice sampling \cite{SliceSampler}.
We then turn to some comparisons between suburban and slice sampling.

\begin{figure}[t!]%
\centering
\includegraphics[
scale = 1.0, trim = 25mm 70mm 0mm 70mm
]%
{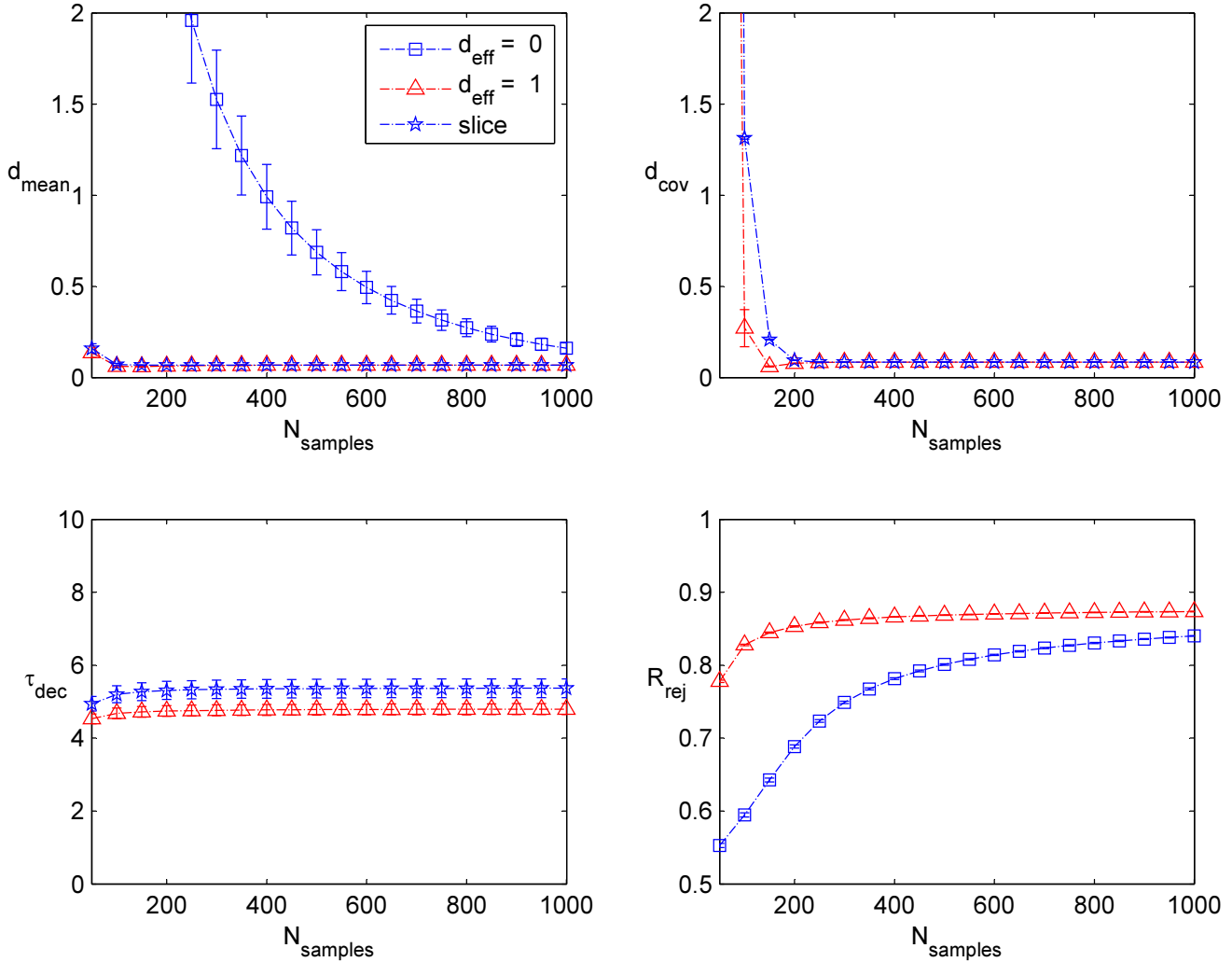}
\caption{Plot of the convergence to the true mean, covariance matrix, as well
as the integrated auto-correlation time and rejection rate for the stringlike $d_{eff} = 1$ suburban sampler
with a 2d grid topology, the parallel MH sampler and parallel slice sampler. For the suburban and MH sampler we take
the brane tension fixed to be $\beta = 0.01$. For the slice sampler we take an initialization width of $1$.
We sample from the same random landscape mixture model with random seed
$40$ discussed in section \ref{sec:LANDSCAPE}. We again perform $T = 100$ independent trials for each choice of hyperparameters.
Performance of the stringlike sampler is comparable to that of the slice sampler.}
\label{LandRand40SliceVsMh}%
\end{figure}

\begin{figure}[t!]%
\centering
\includegraphics[
scale = 1.0, trim = 25mm 70mm 0mm 70mm
]%
{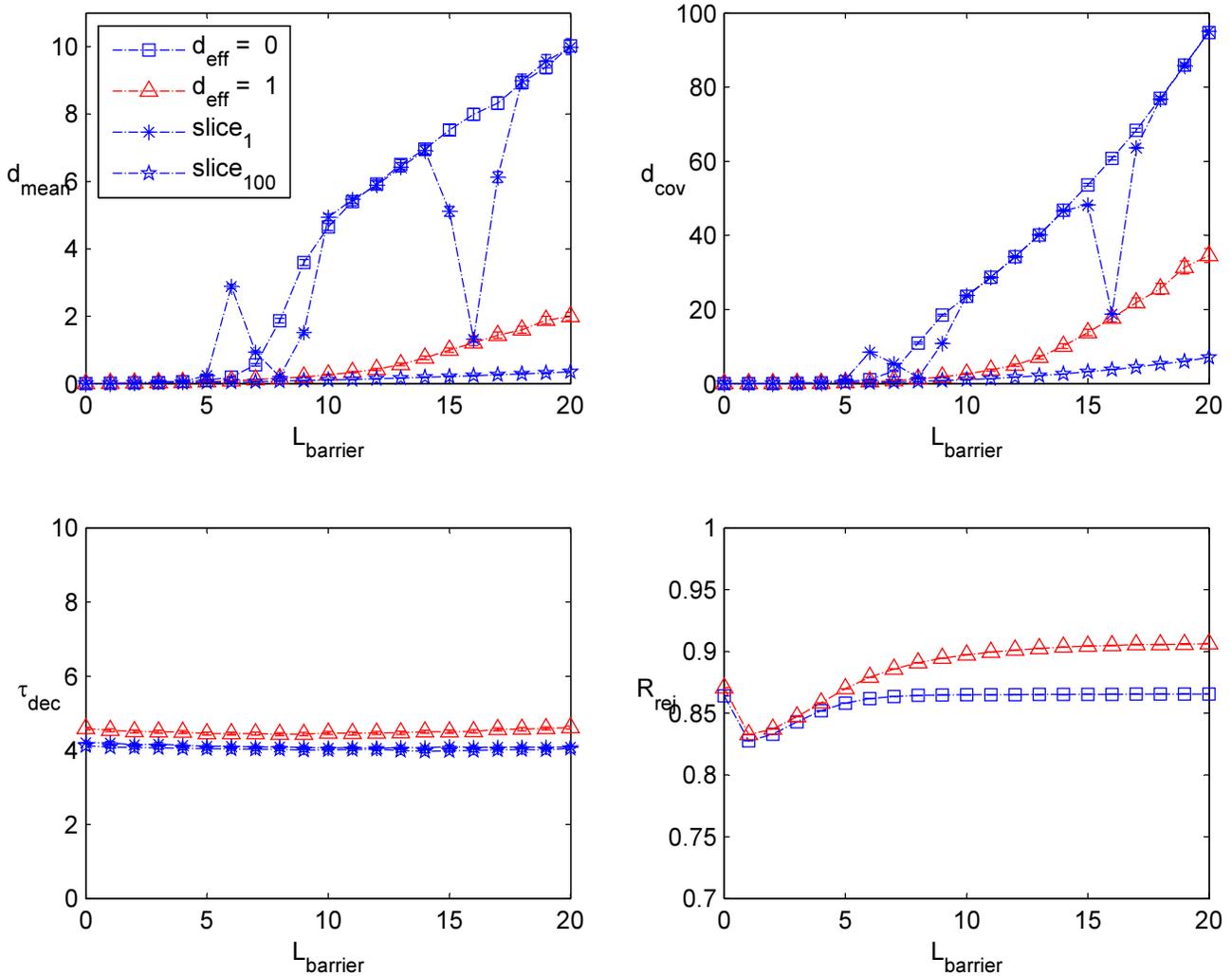}%
\caption{Plot of the convergence to the true mean, covariance matrix, as well
as the integrated auto-correlation time and rejection rate for the suburban
sampler, parallel MH sampler, and two choices of slice sampler with different initialization widths ($1$ and $100$).
We again perform $T = 100$ independent trials for each choice of hyperparameters.
The target distribution is a $D=2$ two component mixture model
where the centers are separated along the x-axis so that the total distance
between the two is $2L_{\text{barrier}}$.}%
\label{2dfreebarrierplotmetNEW}%
\end{figure}

We focus on slice within Gibbs sampling. That is, we sweep over each dimension of the target distribution, treating it
as a univariate distribution on which we apply slice sampling. The particular form of slice sampling we take involves a ``doubling step'' to
increase the size of the horizontal slice. We now proceed to some further details of our use of this algorithm.

Recall that in a one-dimensional slice sampler of a distribution
$\pi(x)$, we pick a point $x_{\ast}$ on the $x$-axis, and then extract the value $\pi(x_{\ast})$. Next, we pick a random value on the $y$-axis
in the interval $(0 , \pi(x_{\ast}))$ using the uniform distribution. At the next step, we introduce an interval $(x_L,x_R)$
of width $w$ containing $x_{\ast}$ with randomly chosen center. If $\pi(x_L) > \pi(x)$ or $\pi(x_R)$, we continue to double the size of the interval by moving the left or right side of the interval further out (randomly picking one of the two sides) until $\pi(x_L) < \pi(x_{\ast})$ and $\pi(x_R) < \pi(x_{\ast})$.

Next, we proceed to pick a new point $x^{\prime}_{\ast}$
on the $x$-axis using a shrinking procedure. First,
designate by $x_{M}$ the midpoint of $(x_L , x_R)$. Draw
a new value $x^{\prime}_{\ast}$ using the uniform distribution
on $(x_L , x_R)$. We reject the new value if $x_{\ast}$ and $x_{\ast}^{\prime}$
are not on opposite sides of the midpoint. If they are on opposite sides of the midpoint,
then we only accept the new value if $\pi(x_{\ast}^{\prime}) > \pi(x_{L})$ and $\pi(x_{\ast}^{\prime}) > \pi(x_{R})$.
If a sample is rejected, we shrink the interval by setting $x_{R}$ to $x_{M}$ when $x_{\ast}^{\prime} < x_{M}$, and otherwise
we move $x_{L}$ to $x_{M}$. The above steps are then repeated until a new sample is accepted. Finally, we repeat all
these steps.

In practice, we mainly use the default implementation in \texttt{Dimple} so that the initial size of the $x$-axis width is an interval of length one containing $x_{\ast}$, and the maximum number of doublings is $10$. In some cases, especially for the free energy barrier tests, this choice of  initialization width can lead to erratic behavior of the sampler. In this case, we find that taking an initial width of $100$ (i.e., much bigger than the size of the separation of the local high density regions) leads to better performance.

Now, as we have already mentioned in section \ref{sec:EXPERIMENT}, it is subtle to directly compare the suburban and slice samplers, since in the former, there is a clear accept/reject choice, while in the latter, everything boils down to the overall size of the intervals and the halting of the ``stepping out'' and ``stepping in'' loops. In practice we find that when we collect some fixed number $N$ of samples, the slice sampler typically makes several more queries to the target distribution compared with the suburban sampler, roughly a factor of $\sim 5 - 10$. Though this makes a direct comparison of the two algorithms less straightforward, we include the results of these tests as a simple way to gauge performance.

To compare the relative performance, we mainly focus on the suburban
sampler obtained from a 2d membrane grid topology, but with effective
dimension $d_{\mathrm{eff}} = 1$. We also focus on the case with brane
tension $\beta = 0.01$.

As a first example, we return to the case of the random landscape model with random seed $40$ studied in section \ref{sec:LANDSCAPE}. For illustrative purposes, in figure \ref{LandRand40SliceVsMh} we compare the tuned suburban sampler with parallel slice samplers. By inspection, we see that we have comparable mixing rates and convergence. We also compare with parallel MH which fares much worse.

As a second example, we consider again the free energy barrier test studied in section \ref{sec:FREEBALL}. Here, we again focus on the suburban sampler since it has the fastest mixing rate. Here, we observe a curious feature: For parallel slice samplers with an initialization width of $1$, we observe erratic behavior in the behavior of the sampler. This appears to be due to the detailed balance requirement associated with the doubling procedure. Indeed, we find that there is no erratic behavior when we take a larger initialization width of $100$ for the slice sampler. Figure \ref{2dfreebarrierplotmetNEW} displays the relative performance of the different samplers for the same 2D free energy barrier test.

\newpage

\bibliographystyle{titleutphys}
\bibliography{Suburban}

\providecommand{\href}[2]{#2}\begingroup\raggedright\begin{thebibliography}{10}

\bibitem{Metropolis:1953am}
N.~Metropolis, A.~W. Rosenbluth, M.~N. Rosenbluth, A.~H. Teller, and E.~Teller,
  ``{Equation of State Calculations by Fast Computing Machines},''
\href{http://dx.doi.org/10.1063/1.1699114}{{\em J. Chem. Phys.} {\bfseries 21}
  (1953) 1087--1092}.

\bibitem{Hastings:1970}
W.~K. Hastings, ``{Monte Carlo Sampling Methods Using Markov Chains and their
  Applications},'' {\em Biometrika} {\bfseries 57} no.~1, (1970) 97--109.

\bibitem{Heckman:2013kza}
J.~J. Heckman, ``{Statistical Inference and String Theory},''
  \href{http://dx.doi.org/10.1142/S0217751X15501602}{{\em Int. J. Mod. Phys.}
  {\bfseries A30} no.~26, (2015) 1550160},
\href{http://arxiv.org/abs/1305.3621}{{\ttfamily arXiv:1305.3621 [hep-th]}}.

\bibitem{SwendsenWang}
R.~H. Swendsen and J.-S. Wang, ``{Replica Monte Carlo Simulation of
  Spin-Glasses},'' {\em Phys. Rev. Lett.} {\bfseries 57} no.~21, (1986)
  2607--2609.

\bibitem{Geyer}
C.~J. Geyer, ``Markov Chain Monte Carlo Maximum Likelihood,'' in {\em Computing
  Science and Statistics: Proceedings of the 23rd Symposium on the Interface},
  E.~M. Keramidas, ed., pp.~156--163.
\newblock Interface Foundation, 1991.

\bibitem{AdaptiveDirectSamp}
W.~R. Gilks, G.~O. Roberts, and E.~I. George, ``{Adaptive Direction
  Sampling},'' {\em Journal of the Royal Statistical Society. Series D (The
  Statistician)} {\bfseries 43} no.~1, (1997) 179--189.

\bibitem{ParallelTempering}
D.~J. Earl and M.~W. Deem, ``{Parallel tempering: Theory, applications, and new
  perspectives},'' {\em Phys. Chem. Chem. Phys.} {\bfseries 7} (2005)
  3910--3916.

\bibitem{NealEnsemble}
R.~M. Neal, ``{MCMC Using Ensembles of States for Problems with Fast and Slow
  Variables such as Gaussian Process Regression},''
  \href{http://arxiv.org/abs/1101.0387}{{\ttfamily arXiv:1101.0387 [stat]}}.

\bibitem{AffineEnsemble}
J.~Goodman and J.~Weare, ``{Ensemble Samplers with Affine Invariance},'' {\em
  Comm. in Appl. Math. and Comp. Sci.} {\bfseries 5} no.~1, (2010) 65--80.

\bibitem{Nishihara}
R.~Nishihara, I.~Murray, and R.~P. Adams, ``Parallel MCMC with Generalized
  Elliptical Slice Sampling,'' {\em J. Mach. Learn. Res.} {\bfseries 15} no.~1,
  (Jan., 2014) 2087--2112.
  \url{http://dl.acm.org/citation.cfm?id=2627435.2670318}.

\bibitem{Exchanging}
K.~{Hukushima} and K.~{Nemoto}, ``{Exchange Monte Carlo Method and Application
  to Spin Glass Simulations},''
  \href{http://dx.doi.org/10.1143/JPSJ.65.1604}{{\em Journal of the Physical
  Society of Japan} {\bfseries 65} (June, 1996) 1604},
  \href{http://arxiv.org/abs/cond-mat/9512035}{{\ttfamily cond-mat/9512035}}.

\bibitem{kou2006}
S.~C. Kou, Q.~Zhou, and W.~H. Wong, ``Equi-energy sampler with applications in
  statistical inference and statistical mechanics,''
  \href{http://dx.doi.org/10.1214/009053606000000515}{{\em Ann. Statist.}
  {\bfseries 34} no.~4, (08, 2006) 1581--1619}.
  \url{http://dx.doi.org/10.1214/009053606000000515}.

\bibitem{Liang00evolutionarymonte}
F.~Liang and W.~H. Wong, ``Evolutionary Monte Carlo: Applications to Cp model
  sampling and change point problem,'' {\em Statistica Sinica} (2000) 317--342.

\bibitem{MultiTry}
J.~S. Liu, F.~Liang, and W.~H. Wong, ``The Multiple-Try Method and Local
  Optimization in Metropolis Sampling,'' {\em Journal of the American
  Statistical Association} {\bfseries 95} no.~449, (2000) 121--134.
  \url{http://www.jstor.org/stable/2669532}.

\bibitem{BalasubramanianGeo}
V.~Balasubramanian, ``{A Geometric Formulation of Occam's Razor For Inference
  of Parametric Distributions},''
\href{http://arxiv.org/abs/adap-org/9601001}{{\ttfamily
  arXiv:adap-org/9601001}}.

\bibitem{Balasubramanian:1996bn}
V.~Balasubramanian, ``{Statistical Inference, Occam's Razor and Statistical
  Mechanics on the Space of Probability Distributions},'' {\em Neural Comp.}
  {\bfseries 9(2)} (1997) 349--368,
\href{http://arxiv.org/abs/cond-mat/9601030}{{\ttfamily
  arXiv:cond-mat/9601030}}.

\bibitem{Heckman:2016jud}
J.~J. Heckman, J.~G. Bernstein, and B.~Vigoda, ``{MCMC with Strings and Branes:
  The Suburban Algorithm},''
\href{http://arxiv.org/abs/1605.06122}{{\ttfamily arXiv:1605.06122 [stat.CO]}}.

\bibitem{Zwiebach:2004tj}
B.~Zwiebach, {\em {A First Course in String Theory}}.
\newblock Cambridge University Press, {Cambridge, UK},
2006.
\newblock

\bibitem{Johnson:2005mqa}
C.~V. Johnson, {\em {D-Branes}}.
\newblock Cambridge University Press, {Cambridge, UK},
2005.
\newblock

\bibitem{Peskin:1995ev}
M.~E. Peskin and D.~V. Schroeder, {\em {An Introduction to quantum field
  theory}}.
\newblock {Addison-Wesley}, {Reading, USA}, 1995.

\bibitem{Wipf:2013vp}
A.~Wipf, ``{Statistical approach to quantum field theory},''
\href{http://dx.doi.org/10.1007/978-3-642-33105-3}{{\em Lect. Notes Phys.}
  {\bfseries 864} (2013) 390}.

\bibitem{Gelman:1997}
A.~Gelman, W.~R. Gilks, and G.~O. Roberts, ``{Weak convergence and optimal
  scaling of random walk Metropolis algorithms},'' {\em Ann. Appl. Prob.}
  {\bfseries 7} no.~1, (1997) 110--120.

\bibitem{DiFrancesco:1997nk}
P.~Di~Francesco, P.~Mathieu, and D.~Senechal,
  \href{http://dx.doi.org/10.1007/978-1-4612-2256-9}{{\em {Conformal Field
  Theory}}}.
\newblock Graduate Texts in Contemporary Physics. Springer-Verlag, New York,
1997.
\newblock

\bibitem{Dimple}
S.~Hershey, J.~Bernstein, B.~Bradley, A.~Schweitzer, N.~Stein, T.~Weber, and
  B.~Vigoda, ``{Accelerating Inference: towards a full Language, Compiler and
  Hardware Stack},'' \href{http://arxiv.org/abs/1212.2991}{{\ttfamily
  arXiv:1212.2991 [cs.SE]}}.

\bibitem{MCHammer}
D.~Foreman-Mackey, D.~W. Hogg, D.~Lang, and J.~Goodman, ``{emcee: The MCMC
  Hammer},'' {\em Pub. of the Ast. Soc. of the Pacific} {\bfseries 125}
  no.~925, (2013) 306--312, \href{http://arxiv.org/abs/1202.3665}{{\ttfamily
  arXiv:1202.3665 [astro-ph.IM]}}.

\bibitem{RafteryLewis}
A.~E. Raftery and S.~M. Lewis, ``{Comment: One Long Run with Diagnostics:
  Implementation Strategies for Markov Chain Monte Carlo},'' {\em Statist.
  Sci.} {\bfseries 7} no.~4, (1992) 493--497.

\bibitem{SliceSampler}
R.~A. Neal, ``{Slice sampling},'' {\em The Ann. of Stat.} {\bfseries 31} no.~3,
  (2003) 705--767.

\bibitem{Herding}
Y.~Chen, M.~Welling, and A.~J. Smola, ``{Super-Samples from Kernel Herding},''
  \href{http://arxiv.org/abs/1203.3472}{{\ttfamily arXiv:1203.3472 [cs.LG]}}.

\end{thebibliography}\endgroup

\end{document}